\documentclass[preprints,article,accept,moreauthors,pdftex,10pt,a4paper]{Definitions/mdpi}
\firstpage{1}
\pubvolume{xx}
\issuenum{1}
\articlenumber{5}
\pubyear{2020}
\copyrightyear{2020}
\history{Received: date; Accepted: date; 
Published: date}

\usepackage{color}
\usepackage{epsfig}
\usepackage{amsmath}
\usepackage{amssymb}
\usepackage{bm}
\usepackage{graphicx}
\graphicspath{ {images/} }
\usepackage{float}

\Title{On Functional Hamilton--Jacobi and Schr\"{o}dinger Equations and Functional Renormalization Group}

\Author{Mikhail G. Ivanov, 
Alexey E. Kalugin, 
Anna A. Ogarkova and 
Stanislav L. Ogarkov*}

\AuthorNames{Mikhail G. Ivanov, Alexey E. Kalugin, Anna A. Ogarkova and Stanislav L. Ogarkov}

\address{%
Moscow Institute of Physics and Technology (MIPT), Institutskiy Pereulok 9, 141701 Dolgoprudny,\\ 
Moscow Region, Russia}

\corres{Correspondence: ogarkovstas@mail.ru}

\abstract{Functional Hamilton--Jacobi (HJ) equation, which is the central equation of the holographic renormalization group (HRG), functional Schr\"{o}dinger equation, and generalized Wilson--Polchinski (WP) equation, which is the central equation of the functional renormalization group (FRG), are considered. These equations are formulated in $D$-dimensional coordinate and abstract (formal) spaces. Instead of extra coordinates or an FRG scale, a ``holographic'' scalar field $\varLambda$ is introduced. The extra coordinate (or scale) is obtained as the amplitude of delta-field or constant-field configurations of $\varLambda$. For all the functional equations above a rigorous derivation of corresponding integro-differential equation hierarchies for Green functions (GFs) as well as the integration formula for functionals are given. An advantage of the HJ hierarchy compared to Schr\"{o}dinger or WP hierarchies is that the HJ hierarchy splits into independent equations. Using the integration formula, the functional (arbitrary configuration of $\varLambda$) solution for the translation-invariant two-particle GF is obtained. For the delta-field and the constant-field configurations of $\varLambda$ this solution is studied in detail. Separable solution for two-particle GF is briefly discussed. Then, rigorous derivation of the quantum HJ and the continuity functional equations from the functional Schr\"{o}dinger equation as well as the semiclassical approximation are given. An iterative procedure for solving the functional Schr\"{o}dinger equation is suggested. Translation-invariant solutions for various GFs (both hierarchies) on delta-field configuration of $\varLambda$ are obtained. In context of continuity equation and open quantum field systems an optical potential is briefly discussed. Modes coarse graining growth functional for WP action (WP functional) is analyzed. Based on this analysis, an approximation scheme is proposed for the generalized WP equation. With an optimized (Litim) regulator translation-invariant solutions for two-particle and four-particle amputated GFs from approximated WP hierarchy are found analytically. For $\varLambda=0$ these solutions are monotonic in each of the momentum variables.}


\usepackage{comment}

\begin{document}

\section{Introduction}

The functional (path) integral formalism underlies many modern theories from various fields of science \cite{zinn1989field,vasil2004field,PopovBook1976,Kleinert,Mosel,Simon,ivanov1993quark,nedelko1995oscillator,brydges1999review,rebenko1988review,efimov1977cmp,efimov1979cmp,efimov2014scattering,efimov2015particle}. For this reason, calculation of functional integrals is an important mathematical problem \cite{Mazzucchi2008,Mazzucchi2009,DeWitt-Morette,Montvay,Johnson,Steiner,Shavgulidze2015,Ogarkov2019I,Ogarkov2019II}. Since a direct calculation is sometimes difficult, one can use the fact that a functional integral is in the general case a solution to some functional equation. In particular, it could be a solution to a partial differential equation, but in general, we have an equation containing variational derivatives. The solution to the latter can be constructed in the form of a series. The simplest series is the functional Taylor series. This ansatz leads us to an infinite system of integro-differential equations (hierarchy or chain) that are coupled to each other in terms of solutions (corresponding GFs family). As an example: The equation for the two-particle GF contains also the four-particle GF in special kinematics. For simplicity we assume that all the GFs of odd order identically equal to zero. This means that we are looking for an even solution to the functional equation. Such a solution exists, for example, in the absence of spontaneous symmetry breaking in the system. The simplicity of this case is that the hierarchy for GFs includes the minimum number of possible terms. Next, the equation for the four-particle GF contains the six-particle GF (in special kinematics) and so on. This is called ``$n,\,n+2$ problem'', and this problem is the main difficulty for functional formulation of QFT (in terms of Dyson--Schwinger and Schwinger--Tomonaga functional equations) and FRG method (in terms of Wilson--Polchinski and Wetterich--Morris functional equations) \cite{kopbarsch,wipf2012statistical,rosten2012fundamentals,igarashi2009realization,efimov1977nonlocal,efimov1985problems,petrina1970ksequations,rebenko1972qed,fradkin2007selected}. Thus, an approximate solution of the hierarchy is based, for example, on the truncation of the hierarchy in a given order.

In comparison with all the examples given, the functional HJ equation has a remarkable property: The hierarchy generated by this equation is decoupled (doesn't contain ``$n,\,n+2$ problem'') \cite{lizana2016holographic}. In this paper, we consider the functional HJ equation first of all. Let us note that this equation is popular today, since it is the basic equation of the HRG method \cite{lizana2016holographic,akhmedov1998remark,de2000holographic,verlinde2000rg,fukuma2003holographic,akhmedov2003notes,akhmedov2011hints,heemskerk2011holographic}. In turn, the HRG is the subject of intensive research, since it is associated with the concept of AdS/CFT correspondence and Holographic Principle \cite{maldacena1999large,witten1998anti,gubser1998gauge,polchinski1998strings}. However, there is still no rigorous derivation of the HRG hierarchy in the literature. Part of our paper aims to fill this gap. 

In the light of holographic applications, we consider the functional HJ equation in terms of an additional ``holographic'' scalar field $\varLambda$ (in the spirit of the paper \cite{doplicher2004generalized}). This field makes the derivation of the HJ hierarchy rigorous and general. Then assuming different configurations of $\varLambda$, it is possible to investigate a number of questions of holography, etc. Further, since the Hamilton functional is a special case of \emph{the modes coarse graining growth functional}, containing only the first variational derivative of the solution (it doesn't contain the second variational derivative of the solution, etc.), it is called the ``holographic'' or ``geometric'' coarse graining growth functional. At the same time, we illustrate the theory in the coordinate representation as well as in the condensed notation (abstract space). This makes it possible to develop geometric intuition regarding the functional HJ equation and HJ hierarchy. Finally, the functional satisfying the HJ equation as well as the corresponding GFs are called holographic functional and holographic GFs, respectively.

Using the first variational derivative of functional integration formula (Newton--Leibniz or NL type formula), we find the translation-invariant solution for the two-particle GF. The derivation of NL type formula is interesting by itself for the following reason: This derivation is similar to the analogous one in case of the FRG flow equations. In two limiting cases (delta-field and constant-field configurations of $\varLambda$), the Riccati equation is obtained for the two-particle GF. Of course, this doesn't prove that the Riccati equation can be obtained for any configuration of $\varLambda$ from the corresponding space of fields, but it allows us to assert the existence of such a space. Concerning the latter equation, we briefly discuss special Riccati equation and self-similar Riccati equation. These equations were obtained in many papers on HRG \cite{lizana2016holographic,akhmedov1998remark,de2000holographic,verlinde2000rg,fukuma2003holographic,akhmedov2003notes,akhmedov2011hints,heemskerk2011holographic}. At the end of the section on the functional HJ equation, we present the separable solution for the two-particle GF on delta-field configuration of $\varLambda$.

The fundamental equation of quantum mechanics (QM) is the Schr\"{o}dinger equation. From this equation, in particular, a pair of equations is derived: The continuity equation and the quantum HJ equation. The latter contains a quantum correction to the classical potential, which can be neglected in the semiclassical approximation. This approach is an alternative to the Wentzel--Kramers--Brillouin approximation and for this reason is one of the semiclassical methods in QM. In our paper, we generalize this derivation to the case of the functional Schr\"{o}dinger equation. Let us note that this equation is similar to the functional diffusion equation. However, the functional Schr\"{o}dinger equation differs significantly due to the ``harmless'' unit imaginary number. We give the rigorous derivation for the pair of hierarchies for the functional continuity equation and the quantum functional HJ equation. It is shown that the functional corresponding to the ``quantum correction'' to the potential in the semiclassical approximation is equal to zero. This is how we arrive at the functional semiclassical approximation. Also, this approach allows us to formulate an iterative procedure for solving the functional Schr\"{o}dinger equation. Let us note that the last one contains the  $n,\,n+2$ problem. Therefore, finding a solution to the functional Schr\"{o}dinger equation without the semiclassical approximation is a rather difficult task.

We study in detail translation-invariant pair of hierarchies (delta-field configuration of $\varLambda$ is assumed): Equations for the vacuum holographic mean (zero-particle holographic GF), two-particle and four-particle holographic GFs, as well as for the vacuum mean and two-particle GF, corresponding to the continuity functional equation. The translation invariance of space (spacetime) leads to the momentum (momentum-energy) conservation law in theory, for example, in Feynman diagrams. \emph{This symmetry is one of the basic symmetries of quantum field theory.} In case of the vacuum mean from continuity functional equation hierarchy, the infinite quantity $(2\pi)^{D}\delta^{(D)}\left(0\right)$ appears. To subtract this quantity (regularize the equation) several strategies can be chosen. We suggest the strategy based on the optical potential as it seems most natural strategy. Such a potential, being complex-valued, contains an imaginary part $\mathcal{W}$. The latter can be split into singular part, absorbing divergences, and regular part, corresponding to the openness of the quantum field system under consideration. The optical potential is well known in QM, where it allows one to transform the exact many-particle Schr\"{o}dinger equation into the effective one-particle one, which is used in various nuclear physics models.

Thus, we come to the important conclusion that the translation-invariant functional Schr\"{o}dinger equation describes an open quantum field system. Let us note separately that the fact that eigenvalues are real follows from the fact that the Hamiltonian is Hermitian. The converse is not true in general. Therefore, one can investigate a separate problem: The construction of a non-Hermitian (functional) Hamiltonian with real eigenvalues (values of energy). However, we don't expect that the functional Schr\"{o}dinger equation in QFT is fundamental. The fundamental equations in QFT are Dyson--Schwinger and Schwinger--Tomonaga functional equations \cite{bogoliubov1980introduction,bogolyubov1983quantum,bogolyubov1990GeneralQFT,Ogarkov2017}. As we noted above, both equations contain the $n,\,n+2$ problem. 

The FRG method is based on a number of functional equations known in QFT, Theory of Critical Phenomena and Stochastic Theory of Turbulence. Probably the most famous are Wilson--Polchinski and Wetterich--Morris functional equations \cite{kopbarsch,wipf2012statistical,rosten2012fundamentals,igarashi2009realization}. WP equation can be derived from the linear functional diffusion equation for the scattering matrix and thus contains the $n,\,n+2$ problem. However, as described in detail in the review \cite{rosten2012fundamentals}, there is an infinite class of the modes coarse graining growth functionals that lead to the WP type equations (generalized WP equation). We assume a subclass of such functionals in our paper. This subclass has a ``geometric nature'' (functionals which depend only on the functional momentum and don't depend on derivatives of the momentum). The latter is most transparent in condensed notation (abstract space) \cite{kopbarsch}. The validity of this assumption is investigated according to the general definition of the modes coarse graining growth functional in terms of the functional integral. Let us note that this definition is the starting point for the Higher Regulators Theory. Also, this definition contains the Dirac delta-functional, which makes it rather complicated for direct calculation of corresponding functional integral.

Despite the fact that WP equation doesn't contain an unit imaginary number, it can also be divided into classical and quantum parts. Therefore, it is possible to implement a certain analogue of functional semiclassical approximation. However, this semiclassics differs from that for the functional Schr\"{o}dinger equation. Also in the WP case, the knowledge of the system's classical action (as a boundary condition) is required. The latter is used for the approximation of the WP equation quantum part. In this paper we obtain the solution for the approximated WP hierarchy. More specifically, we find translation-invariant two-particle and four-particle amputated GFs analytically (delta-field $\varLambda$ is assumed). As an example: The equation for the two-particle amputated GF is the special Riccati equation. The solution to the latter is a combination of modified Bessel functions. For the four-particle amputated GF, the equation turns out to be linear. In all calculations we use the optimized (Litim) regulator \cite{kopbarsch,wipf2012statistical}. We also discuss the physical ($\varLambda=0$) two-particle and four-particle amputated GFs for different values of the coupling constant in the the polynomial $\varphi^{4}$ model in three-dimensional space with a quadratic massless propagator. Let us note here, that the obtained approximate solution contains information about critical dimensions, corresponding to the emergence of conformal field theories (CFTs) \cite{zinn1989field,vasil2004field}.

\emph{This paper has the following structure:} In section $2$ we investigate the functional Hamilton–Jacobi equation. Section $3$ is devoted to the functional Schr\"{o}dinger equation. Section $4$ is devoted to the general Wilson--Polchinski equation. In the Conclusions, section $5$, we give a final discussion of all the results obtained in the paper and highlight further possible areas of research. \emph{The main results obtained by the authors and presented in this paper are:} rigorous derivation of the functional HJ equation hierarchy, solution of the first equations of the hierarchy for an arbitrary field $\varLambda$, rigorous derivation of the functional Schr\"{o}dinger equation hierarchy (quantum HJ and continuity functional equations hierarchies,
semiclassical solution of the first equations of the hierarchies (with optical potential), rigorous derivation of the WP functional equation from the functional master equation (the equation containing the modes coarse graining growth functional), semiclassical solution of the first equations of the WP functional equation hierarchy (with Litim regulator), rigorous derivation of conditions when the modes coarse graining growth functional is geometric.

\section{Functional Hamilton--Jacobi Equation and its Hierarchy}

In this section, we consider the functional HJ equation. We draw parallels between the coordinate representation and condensed notation in order to develop intuition about the latter: The presentation of the functional Schr\"{o}dinger equation will be in terms of condensed notation until a certain point. According to the Introduction section, we use the terminology ``holographic'' field, ``holographic'' GFs, etc. This terminology is caused by applications of the functional HJ equation in the HRG method \cite{lizana2016holographic,akhmedov1998remark,de2000holographic,verlinde2000rg,fukuma2003holographic,akhmedov2003notes,akhmedov2011hints,heemskerk2011holographic}. However, the equation itself appeared much earlier than the HRG method, and the area of its application is much wider than the HRG. For this reason, a reader unfamiliar with the HRG can omit this terminology without losing the meaning of what he reads in this section.

\subsection{Functional Hamilton--Jacobi Equation}

Let us introduce the following notation: $\varphi$ is the scalar field which is the function of the spatial coordinate $x$, $\varLambda$ is the holographic (scalar) field which is the function of the spatial coordinate $z$, $\varpi$ is the functional momentum which is the function of the spatial coordinate $x$ and the functional of the filed $\varLambda$. \emph{Convention: The dependence of the functional on the function is denoted with square brackets, while the dependence of the function on its argument is denoted with parentheses.} The functional HJ equation for the holographic functional $\mathcal{S}$ in the coordinate representation is given by the following deceptively simple expression ($x,y,z\in\mathbb{R}^{D}$ although in general, $z$ can refer to a \emph{different} space with different $D$):
\begin{equation}
    \frac{\delta\mathcal{S}\left[\varLambda, \varphi\right]}{\delta\varLambda\left(z\right)}=
    \mathcal{H}\left[\varLambda,\varphi,\varpi\right]
    \left(z\right),\quad \varpi\left(x\right)=\frac{\delta\mathcal{S}
    \left[\varLambda,\varphi\right]}
    {\delta\varphi\left(x\right)}.
\label{HRG_Equation_CR}
\end{equation}

The functional HJ equation in the condensed notation is ($\alpha,\mu,\nu$ are abstract or formal indices, which can include continuous variables (spatial coordinate, time, momentum, frequency), discrete variables (spin, isospin, flavor), abstract manifold variables, etc.):
\begin{equation}
    \frac{\delta\mathcal{S}\left[\varLambda, \varphi\right]}{\delta\varLambda_{\alpha}}=
    \mathcal{H}_{\alpha}
    \left[\varLambda,\varphi,\varpi\right],\quad
    \varpi_{\mu}=\frac{\delta\mathcal{S}
    \left[\varLambda,\varphi\right]}
    {\delta\varphi_{\mu}}.
\label{HRG_Equation_CN}
\end{equation}

The holographic functional $\mathcal{S}$ is a solution to the equation, the Hamilton functional $\mathcal{H}$ is the holographic or geometric growth functional. The latter can be expanded in the functional Taylor series over the field $\varpi$. Corresponding expression in the coordinate representation is:
\begin{equation}
\begin{split}
    &\mathcal{H}
    \left[\varLambda,\varphi,\varpi\right]
    \left(z\right)=
    \mathcal{H}^{(0)}
    \left[\varLambda,\varphi\right]
    \left(z\right)+\\
    &+\sum\limits_{n=1}^{\infty}
    \frac{1}{n!}
    \int d^D x_{1}\ldots\int d^D x_{n}
    \mathcal{H}^{(n)}
    \left[\varLambda,\varphi\right]
    \left(z;x_{1},\ldots,x_{n}\right)
    \varpi\left(x_{1}\right)\ldots
    \varpi\left(x_{n}\right).
    \label{Hamilton_Functional_TS_CR}
\end{split}
\end{equation}

The expression in the condensed notation reads:
\begin{equation}
    \mathcal{H}_{\alpha}
    \left[\varLambda,\varphi,\varpi\right]=
    \mathcal{H}^{(0)}_{\alpha}
    \left[\varLambda,\varphi\right]+
    \sum\limits_{n=1}^{\infty}
    \frac{1}{n!}\int_{\mu_{1}}\ldots
    \int_{\mu_{n}}
    \mathcal{H}^{(n)}_{\alpha,
    \mu_{1}\ldots\mu_{n}}
    \left[\varLambda,\varphi\right]
    \varpi_{\mu_{1}}\ldots\varpi_{\mu_{n}}.
    \label{Hamilton_Functional_TS_CN}
\end{equation}

To mimic classical physics (analytical mechanics), we restrict ourselves to the zero and second terms (the symmetry $\varpi\rightarrow -\varpi$ is assumed, hence $\mathcal{H}^{(1)}=0$) in the expansion (\ref{Hamilton_Functional_TS_CR}). Thus, we arrive at the following equation in the coordinate representation ($\mathcal{H}^{(0)}= \mathcal{U}$ and $\mathcal{H}^{(2)}= \mathcal{T}$):
\begin{equation}
    \frac{\delta\mathcal{S}\left[\varLambda, \varphi\right]}{\delta\varLambda\left(z\right)}=
    \mathcal{U}\left[\varLambda,\varphi\right]\left(z\right)+\frac{1}{2}\int d^D x_{1}\int d^D x_{2} \mathcal{T}\left[\varLambda, \varphi\right]\left(z;x_{1},x_{2}\right)
    \frac{\delta \mathcal{S}\left[\varLambda,\varphi\right]}
    {\delta\varphi\left(x_{1}\right)}
    \frac{\delta \mathcal{S}\left[\varLambda,\varphi\right]}
    {\delta\varphi\left(x_{2}\right)}.
    \label{Functional_HJ_Equation_CR}
\end{equation}

Restricting ourselves to the zero and second terms (the symmetry $\varpi\rightarrow -\varpi$ is assumed, which means the absence of a term linear in functional momentum in the Hamilton functional -- a common situation in classical physics when there are no dissipative forces in the system) in the expansion (\ref{Hamilton_Functional_TS_CN}), we arrive at the following equation in the condensed notation:
\begin{equation}
\begin{split}
    \frac{\delta\mathcal{S}\left[\varLambda, \varphi\right]}{\delta\varLambda_{\alpha}}=
    \mathcal{U}_{\alpha}\left[\varLambda, \varphi\right]+\frac{1}{2}\int_{\mu_{1}}
    \int_{\mu_{2}}\mathcal{T}_{\alpha,\mu_{1}\mu_{2}}
    \left[\varLambda,\varphi\right]\frac{\delta \mathcal{S}\left[\varLambda,\varphi\right]}{\delta\varphi_{\mu_{1}}}\frac{\delta \mathcal{S}\left[\varLambda,\varphi\right]}{\delta\varphi_{\mu_{2}}}.
    \label{Functional_HJ_Equation_CN}
\end{split}
\end{equation}

Let us consider a solution of the equation (\ref{Functional_HJ_Equation_CR}) in the form of the functional Taylor series over the field $\varphi$ in the coordinate representation:
\begin{equation}
    \mathcal{S}\left[\varLambda, \varphi\right]=
    \mathcal{S}^{(0)}\left[\varLambda\right]+
    \sum_{n=1}^{\infty}\frac{1}{n!}
    \int d^D y_{1}\ldots\int d^D y_{n}\,
    \mathcal{S}^{(n)}\left[\varLambda\right]
    \left(y_{1},\ldots,y_{n}\right)
    \varphi\left(y_{1}\right)\ldots
    \varphi\left(y_{n}\right),
    \label{HJ_Solution_CR}
\end{equation}
where the functions $\mathcal{S}^{(n)}$, $n\in\mathbb{N}_{0}$ are the holographic GFs family, that is the solution to the HJ hierarchy. In the condensed notation the expression (\ref{HJ_Solution_CR}) reads:
\begin{equation}
    \mathcal{S}\left[\varLambda,\varphi\right]=
    \mathcal{S}^{(0)}\left[\varLambda\right]+
    \sum_{n=1}^{\infty}\frac{1}{n!}
    \int_{\nu_{1}}\ldots\int_{\nu_{n}}
    \mathcal{S}^{(n)}_{\nu_{1}\ldots\nu_{n}}
    \left[\varLambda\right]
    \varphi_{\nu_{1}}\ldots
    \varphi_{\nu_{n}}.
    \label{HJ_Solution_CN}
\end{equation}

To understand the further derivation, consider the variational derivatives of $\mathcal{S}$ over the field $\varphi$. The first variational derivative is:
\begin{equation}
    \frac{\delta\mathcal{S}
    \left[\varLambda,\varphi\right]}
    {\delta\varphi\left(x\right)}=\int d^D y\, \mathcal{S}^{(2)}\left[\varLambda\right]
    \left(x,y\right)\varphi\left(y\right)+
    \mathcal{O}\left[\varphi^{3}\right],\quad
    \frac{\delta\mathcal{S}
    \left[\varLambda,\varphi\right]}
    {\delta\varphi_{\mu}}=\int_{\nu} \mathcal{S}^{(2)}_{\mu\nu}\left[\varLambda\right]
    \varphi_{\nu}+\mathcal{O}\left[\varphi^{3}\right].
\end{equation}

The second variational derivative is:
\begin{equation}
    \frac{\delta^{2}\mathcal{S}
    \left[\varLambda,\varphi\right]}
    {\delta\varphi\left(x\right)
    \delta\varphi\left(y\right)}= \mathcal{S}^{(2)}\left[\varLambda\right]
    \left(x,y\right)+
    \mathcal{O}\left[\varphi^{2}\right],\quad
    \frac{\delta^{2}\mathcal{S}
    \left[\varLambda,\varphi\right]}
    {\delta\varphi_{\mu}
    \delta\varphi_{\nu}}=
    \mathcal{S}^{(2)}_{\mu\nu}
    \left[\varLambda\right]+
    \mathcal{O}\left[\varphi^{2}\right].
\end{equation}

Thus, we see that the easiest way to obtain the HJ hierarchy is to differentiate equations (\ref{Functional_HJ_Equation_CR}) or (\ref{Functional_HJ_Equation_CN}) directly. This is done in the next subsection.

\subsection{Two-Particle Green Function Equation}

Let us derive the equation for the two-particle holographic GF. This function is the most important of all GFs: Higher-order holographic GFs can be obtained from linear equations, where the two-particle holographic GF gives the coefficients of the equations. First, we take the second variational derivative of the HJ equation (\ref{Functional_HJ_Equation_CR}):
\begin{equation}
\begin{split}
    &\frac{\delta}{\delta\varLambda\left(z\right)}
    \frac{\delta^{2}\mathcal{S}
    \left[\varLambda,\varphi\right]}
    {\delta\varphi\left(y_{1}\right)
    \delta\varphi\left(y_{2}\right)}=
    \frac{\delta^{2}\mathcal{U}\left[\varLambda,\varphi\right]
    \left(z\right)}{\delta\varphi\left(y_{1}\right)
    \delta\varphi\left(y_{2}\right)}+\\
    &+\int d^D x_{1}\int d^D x_{2} \mathcal{T}\left[\varLambda,\varphi\right]
    \left(z;x_{1},x_{2}\right)
    \frac{\delta^{2}\mathcal{S}
    \left[\varLambda,\varphi\right]\left(z\right)}
    {\delta\varphi\left(x_{1}\right)
    \delta\varphi\left(y_{1}\right)}
    \frac{\delta^{2}\mathcal{S}
    \left[\varLambda,\varphi\right]\left(z\right)}
    {\delta\varphi\left(x_{2}\right)
    \delta\varphi\left(y_{2}\right)}+\ldots
    \label{Derivation_Interm}
\end{split}
\end{equation}

Dots in (\ref{Derivation_Interm}) mean terms containing variational derivatives of the functional $\mathcal{T}$ with respect to field $\varphi$. The definition of $n$-particle holographic GF in terms of variational derivatives in the coordinate representation is \cite{kopbarsch}:
\begin{equation}
    \frac{\delta^{n}\mathcal{S}
    \left[\varLambda,\varphi\right]}
    {\delta\varphi\left(y_{1}\right)\ldots
    \delta\varphi\left(y_{n}\right)}
    \bigg\rvert_{\varphi=0}=\mathcal{S}^{(n)}
    \left[\varLambda\right]
    \left(y_{1},\ldots,y_{n}\right).
    \label{Green_Function_CR}
\end{equation}

The definition of $n$-particle holographic GF in terms of variational derivatives in the condensed notation reads as follows \cite{kopbarsch}:
\begin{equation}
    \frac{\delta^{n}\mathcal{S}
    \left[\varLambda,\varphi\right]}
    {\delta\varphi_{\nu_{1}}\ldots
    \delta\varphi_{\nu_{n}}}
    \bigg\rvert_{\varphi=0}=
    \mathcal{S}^{(n)}_{\nu_{1}\ldots\nu_{n}}
    \left[\varLambda\right].
    \label{Green_Function_CN}
\end{equation}

According to the definition (\ref{Green_Function_CR}), setting the field $\varphi=0$ in the equation (\ref{Derivation_Interm}), we arrive at the following equation for the two-particle holographic GF in the coordinate representation is:
\begin{equation}
\begin{split}
    &\frac{\delta\mathcal{S}^{(2)}
    \left[\varLambda\right]
    \left(y_{1},y_{2}\right)}
    {\delta\varLambda\left(z\right)}=
    \mathcal{U}^{(2)}\left[\varLambda\right]
    \left(z;y_{1},y_{2}\right)+\\
    &+\int d^D x_{1}\int d^D x_{2}
    \mathcal{T}^{(0)}
    \left[\varLambda\right]
    \left(z;x_{1},x_{2}\right)
    \mathcal{S}^{(2)}
    \left[\varLambda\right]
    \left(x_{1},y_{1}\right)
    \mathcal{S}^{(2)}
    \left[\varLambda\right]
    \left(x_{2},y_{2}\right).
    \label{Equation_Two-Particle_GF_CR}
\end{split}
\end{equation}

The equation for two-particle equation for the two-particle holographic GF in the condensed notation reads as follows:
\begin{equation}
    \frac{\delta\mathcal{S}^{(2)}_{\nu_{1}\nu_{2}}
    \left[\varLambda\right]}
    {\delta\varLambda_{\alpha}}=
    \mathcal{U}^{(2)}_{\alpha,\nu_{1}\nu_{2}}
    \left[\varLambda\right]+
    \int_{\mu_{1}}\int_{\mu_{2}}
    \mathcal{T}^{(0)}_{\alpha,\mu_{1}\mu_{2}}
    \left[\varLambda\right]
    \mathcal{S}^{(2)}_{\mu_{1}\nu_{1}}
    \left[\varLambda\right]
    \mathcal{S}^{(2)}_{\mu_{2}\nu_{2}}
    \left[\varLambda\right].
    \label{Equation_Two-Particle_GF_CN}
\end{equation}

Equations (\ref{Equation_Two-Particle_GF_CR})--(\ref{Equation_Two-Particle_GF_CN}) are complicated integro-differential equations. However, the presence of additional symmetries in the system can significantly simplify these equations. In the next subsection, we consider such a case.

\subsection{Translation-Invariant Solution for Green Function on Delta-Field Configuration}

In QFT and Theory of Critical Phenomena, the translation invariance is the most natural 
symmetry \cite{zinn1989field,vasil2004field}. For this reason let us consider translation-invariant problem in the coordinate representation:
\begin{equation}
\begin{split}
    \mathcal{U}^{(2)}\left[\varLambda\right]
    \left(z;y_{1},y_{2}\right)=
    \mathcal{U}\left[\varLambda\right]
    \left(z;y_{1}-y_{2}\right)=
    \int_{k}e^{ik(y_{1}-y_{2})} 
    \mathcal{U}\left[\varLambda\right]
    \left(z;k\right),\\
    \mathcal{T}^{(0)}\left[\varLambda\right]
    \left(z;y_{1},y_{2}\right)=
    \mathcal{T}\left[\varLambda\right]
    \left(z;y_{1}-y_{2}\right)=
    \int_{k}e^{ik(y_{1}-y_{2})} 
    \mathcal{T}\left[\varLambda\right]
    \left(z;k\right).
    \label{Translation-invariant_problem}
\end{split}
\end{equation}

For the Fourier transform, we use compact notation for the $D$-fold integration over the momentum space which is $\mathbb{R}^{D}$:
\begin{equation}
    \int_{k}\equiv\int\frac{d^{D}k}{(2\pi)^{D}}=
    \int\frac{dk_{1}}{2\pi}\ldots
    \int\frac{dk_{D}}{2\pi}
    \Rightarrow\int_{k}
    (2\pi)^{D}\delta^{(D)}\left(k\right)=1.
    \label{Momentum_Compact_Notation}
\end{equation}

Since the symmetry of the solution often preserves the symmetry of the equation coefficients, let us consider translation-invariant two-particle holographic GF in the coordinate representation:
\begin{equation}
    \mathcal{S}^{(2)}\left[\varLambda\right]
    \left(y_{1},y_{2}\right)=
    \mathcal{S}\left[\varLambda\right]
    \left(y_{1}-y_{2}\right)=
    \int_{k}e^{ik(y_{1}-y_{2})} 
    \mathcal{S}\left[\varLambda\right]\left(k\right).
    \label{Translation-invariant_solution}
\end{equation}

Further, the equation (\ref{Equation_Two-Particle_GF_CR}) for two-particle holographic GF for translation-invariant problem (\ref{Translation-invariant_problem}) in coordinate representation is:
\begin{equation}
    \begin{split}
    &\frac{\delta\mathcal{S}\left[\varLambda\right]
    \left(y_{1}-y_{2}\right)}
    {\delta\varLambda\left(z\right)}=
    \mathcal{U}\left[\varLambda\right]
    \left(z;y_{1}-y_{2}\right)+\\
    &+\int d^D x_{1}\int d^D x_{2}
    \mathcal{T}
    \left[\varLambda\right]
    \left(z;x_{1}-x_{2}\right)
    \mathcal{S}\left[\varLambda\right]
    \left(x_{1}-y_{1}\right)
    \mathcal{S}\left[\varLambda\right]
    \left(x_{2}-y_{2}\right).
    \label{Equation_Two-Particle_GF_TIP_CR}
\end{split}
\end{equation}

The equation (\ref{Equation_Two-Particle_GF_TIP_CR}) for two-particle holographic GF for translation-invariant problem in momentum representation reads as follows (equation with ``diagonal'' right-hand side):
\begin{equation}
    \frac{\delta\mathcal{S}\left[\varLambda\right]
    \left(k\right)}
    {\delta\varLambda\left(z\right)}=
    \mathcal{U}\left[\varLambda\right]
    \left(z;k\right)+
    \mathcal{T}\left[\varLambda\right]
    \left(z;k\right)
    \mathcal{S}\left[\varLambda\right]
    \left(-k\right)
    \mathcal{S}\left[\varLambda\right]
    \left(k\right).
    \label{Equation_Two-Particle_GF_TIP_MR}
\end{equation}

In the following subsections, using the integration formula for functionals, we will rewrite the equation (\ref{Equation_Two-Particle_GF_TIP_MR}) without the first variational derivative over $\varLambda$. However, in order to develop some intuition regarding the equation (\ref{Equation_Two-Particle_GF_TIP_MR}), in this subsection we choose another strategy. Let us consider two-particle holographic GF $\mathcal{S}$ as a functional of the field $\varLambda$ and its functional Taylor series over $\varLambda$ in the coordinate representation:
\begin{equation}
    \mathcal{S}\left[\varLambda\right]\left(k\right)=
    \mathcal{S}^{(0)}\left(k\right)+
    \sum_{n=1}^{\infty}\frac{1}{n!}
    \int d^D z_{1}\ldots\int d^D z_{n}\, \mathcal{S}^{(n)}\left(z_{1},\ldots,z_{n};k\right)
    \varLambda\left(z_{1}\right)\ldots
    \varLambda\left(z_{n}\right).
    \label{FTS_Two-Particle_GF_CR}
\end{equation}

Variational derivative of expression (\ref{FTS_Two-Particle_GF_CR}) reads as follows:
\begin{equation}
    \frac{\delta\mathcal{S}
    \left[\varLambda\right]\left(k\right)}
    {\delta\varLambda\left(z\right)}=
    \mathcal{S}^{(1)}\left(z;k\right)+
    \sum_{n=2}^{\infty}\frac{1}{\left(n-1\right)!}
    \int d^D z_{2}\ldots\int d^D z_{n}\, \mathcal{S}^{(n)}\left(z,z_{2},\ldots,z_{n};k\right)
    \varLambda\left(z_{2}\right)\ldots
    \varLambda\left(z_{n}\right).
    \label{Der_FTS_Two-Particle_GF_CR}
\end{equation}

Let's take advantage of the fact that we can choose from different field configurations of $\varLambda$. We introduce delta-field configuration of $\varLambda$ ($w$ is an arbitrary parameter, $D$ is the dimension of the space on which the field $\varLambda$ is defined). From the geometric point of view, we choose the direction in the space of functions. The results obtained with such a choice, qualitatively remain valid for a wide class of functions and spaces of functions:
\begin{equation}
    \varLambda\left(z_{l}\right) = \varLambda\,\delta^{(D)}\left(z_{l}-w\right) 
    \equiv\varLambda_{\delta,w}.
    \label{Delta-field_configuration}
\end{equation}

For the delta-field configuration of $\varLambda$ the expression (\ref{FTS_Two-Particle_GF_CR}) is:
\begin{equation}
    \mathcal{S}\left[\varLambda\right]\left(k\right)
    \big\rvert_{\varLambda_{\delta,w}}=
    \mathcal{S}^{(0)}\left(k\right)+
    \sum_{n=1}^{\infty}\frac{\varLambda^{n}}{n!}\,
    \mathcal{S}^{(n)}\left(w,\ldots,w;k\right)\equiv \mathcal{S}\left(\varLambda;w;k\right).
    \label{FTS_Two-Particle_GF_Delta}
\end{equation}

For the delta-field configuration of $\varLambda$ the expression (\ref{Der_FTS_Two-Particle_GF_CR}) reads as follows:
\begin{equation}
    \frac{\delta\mathcal{S}
    \left[\varLambda\right]\left(k\right)}
    {\delta\varLambda\left(z\right)}
    \bigg\rvert_{\varLambda_{\delta,w}}=
    \mathcal{S}^{(1)}\left(z;k\right)+
    \sum_{n=2}^{\infty}\frac{\varLambda^{n-1}}
    {\left(n-1\right)!}\, \mathcal{S}^{(n)}\left(z,w,\ldots,w;k\right).
    \label{Der_FTS_Two-Particle_GF_Delta}
\end{equation}

The expressions (\ref{FTS_Two-Particle_GF_Delta})--(\ref{Der_FTS_Two-Particle_GF_Delta}) demonstrate the following fact: If we choose delta-field configuration of $\varLambda$ in the equation (\ref{Equation_Two-Particle_GF_TIP_MR}) and then set $z=w$, the equation (\ref{Equation_Two-Particle_GF_TIP_MR}) becomes closed (up to $k\rightarrow -k$ symmetry) with respect to the function $\mathcal{S}\left(\varLambda;w;k\right)$ (\ref{FTS_Two-Particle_GF_Delta}):
\begin{equation}
    \frac{\partial\mathcal{S}
    \left(\varLambda;w;k\right)}
    {\partial\varLambda}=
    \mathcal{U}\left(\varLambda;w;k\right)+
    \mathcal{T}\left(\varLambda;w;k\right)
    \mathcal{S}\left(\varLambda;w;-k\right)
    \mathcal{S}\left(\varLambda;w;k\right).
    \label{Equation_Two-Particle_GF_TIP_Delta}
\end{equation}

If coefficient functions $\mathcal{U}\left(\varLambda;w;k\right)$ and $\mathcal{T}\left(\varLambda;w;k\right)$ are even functions of $k$, we obtain closed with respect to the function $\mathcal{S}\left(\varLambda;w;k\right)$ (\ref{FTS_Two-Particle_GF_Delta}) equation. This means that the corresponding functionals on the left-hand sides of equality (\ref{Translation-invariant_problem}) are even functionals with respect to the permutation of $y_{1}$ and $y_{2}$. The latter is always true, since these functionals are the coefficients of the functional Taylor series. In order to be able to consider odd functionals, the very formulation of the problem should be changed. Thus, we arrive at the following equation:
\begin{equation}
    \frac{\partial\mathcal{S}
    \left(\varLambda;w;k\right)}
    {\partial\varLambda}=
    \mathcal{U}\left(\varLambda;w;k\right)+
    \mathcal{T}\left(\varLambda;w;k\right)
    \mathcal{S}^{2}\left(\varLambda;w;k\right).
    \label{Riccati_Equation_Two-Particle_GF_TIP}
\end{equation}

The equation (\ref{Riccati_Equation_Two-Particle_GF_TIP}) is the Riccati equation. For arbitrary even coefficient functions $\mathcal{U}\left(\varLambda;w;k\right)$ and $\mathcal{T}\left(\varLambda;w;k\right)$, the Riccati equation can't be integrated by quadratures. Some special cases of the Riccati equation arising in the HRG and FRG are given below.

\subsubsection{Special Riccati Equation}

Let us consider the case of the special Riccati equation (the definition of this equation can be found in any handbook on differential equations). To obtain the special Riccati equation, the coefficient functionals $\mathcal{U}\left[\varLambda\right]\left(z;k\right)$ and $\mathcal{T}\left[\varLambda\right]\left(z;k\right)$ should be chosen as follows (with sufficiently arbitrary functions $\mathcal{A}$ and $\mathcal{B}$ and arbitrary parameters $a$ and $b$): 
\begin{equation}
\begin{split}
    \mathcal{T}\left[\varLambda\right]
    \left(z;k\right)=
    &-\left\{\int d^D z'\mathcal{A}^{\frac{1}{a}}\left(z';z;k\right)
    \varLambda\left(z'\right)\right\}^{a},\\
    \mathcal{U}\left[\varLambda\right]
    \left(z;k\right)=
    &+\left\{\int d^D z'\mathcal{B}^{\frac{1}{b}}\left(z';z;k\right)
    \varLambda\left(z'\right)\right\}^{b}.
    \label{Bessel_1}
\end{split}
\end{equation}

For the delta-field configuration of $\varLambda$ and $z=w$ the expression (\ref{Bessel_1}) is:
\begin{equation}
    \mathcal{T}\left(\varLambda;w;k\right)=
    -\mathcal{A}\left(w;w;k\right)
    \varLambda^{a},\quad
    \mathcal{U}\left(\varLambda;w;k\right)=
    +\mathcal{B}\left(w;w;k\right)
    \varLambda^{b}.
    \label{Bessel_2}
\end{equation}

The equation (\ref{Riccati_Equation_Two-Particle_GF_TIP}) becomes a special Riccati equation ($w$ and $k$ dependencies are omitted for compactness of further expressions):
\begin{equation}
    \frac{\partial\mathcal{S}
    \left(\varLambda\right)}
    {\partial\varLambda}=
    \mathcal{B}\varLambda^{b}
    -\mathcal{A}\varLambda^{a}
    \mathcal{S}^{2}\left(\varLambda\right).
    \label{Bessel_3}
\end{equation}

General solution of the special Riccati equation (\ref{Bessel_3}) in terms of modified Bessel functions $I$ and $K$ reads as follows ($\mathcal{C}$ is a ``constant'' of integration -- function that doesn't depend on $\varLambda$):
\begin{equation}
    \mathcal{S}\left(\varLambda\right)=
    \mathcal{B}\varLambda^{b+1}\,
    \frac{\mathcal{C}K_{d}\left(X\right)+
    I_{d}\left(X\right)}
    {\mathcal{C}\tilde{K}_{d}\left(X\right)+
    \tilde{I}_{d}\left(X\right)}.
    \label{Bessel_4}
\end{equation}

In expression (\ref{Bessel_4}) we use the following compact notation:
\begin{equation}
\begin{split}
    \tilde{K}_{d}\left(X\right)&=
    \left(b+1\right)K_{d}\left(X\right)-
    \frac{a+b+2}{2}XK_{d+1}\left(X\right),\\
    \tilde{I}_{d}\left(X\right)&=
    \left(b+1\right)I_{d}\left(X\right)+
    \frac{a+b+2}{2}XI_{d+1}\left(X\right),\\
    X&=\frac{2\sqrt{\mathcal{AB}}
    \varLambda^{\frac{a+b+2}{2}}}
    {a+b+2},\quad d=\frac{b+1}{a+b+2}.
    \label{Bessel_5}
\end{split}
\end{equation}

The solution (\ref{Bessel_5}) is found in the literature in papers on HRG \cite{lizana2016holographic,akhmedov1998remark,de2000holographic,verlinde2000rg,fukuma2003holographic,akhmedov2003notes,akhmedov2011hints,heemskerk2011holographic}. We will also encounter it when solving the equations of the approximate WP hierarchy.

\subsubsection{Self-Similar Riccati Equation}

Let us consider the self-similar Riccati equation case. In this case, the equation (\ref{Riccati_Equation_Two-Particle_GF_TIP}) has a solution $\mathcal{S}\left(\varLambda;w;k\right)$ of the following form (the parameter $a$ is the canonical dimension of the function $\mathcal{S}$ and the function $\widetilde{\mathcal{S}}$ is the dimensionless function of new self-similar variables $\omega$ and $\varkappa$):
\begin{equation}
    \mathcal{S}\left(\varLambda;w;k\right)=
    \varLambda^{a}
    \widetilde{\mathcal{S}}
    \left(\omega\left(\varLambda;w\right)\!;
    \varkappa\left(\varLambda;k\right)\right).
    \label{relation_of_variables}
\end{equation}

The left-hand side of the equation (\ref{Riccati_Equation_Two-Particle_GF_TIP}) -- the total derivative of the function $\mathcal{S}\left(\varLambda;w;k\right)$ over $\varLambda$ can be represented as follows:
\begin{equation}
    \frac{\partial\mathcal{S}
    \left(\varLambda;w;k\right)}{\partial
    \varLambda}=\varLambda^{a-1}
    \left\{a+\varLambda
    \frac{\partial\omega}
    {\partial\varLambda}\frac{\partial}
    {\partial\omega}+\varLambda
    \frac{\partial\varkappa}
    {\partial\varLambda}
    \frac{\partial}
    {\partial\varkappa}\right\}
    \widetilde{\mathcal{S}}
    \left(\omega;\varkappa\right).
    \label{compound_derivative}
\end{equation}

By definition, self-similar solution $\widetilde{\mathcal{S}}$ doesn't explicitly depend on $\varLambda$, therefore, for such a solution, the equation (\ref{Riccati_Equation_Two-Particle_GF_TIP}) can be rewritten in the following form:
\begin{equation}
    \left\{a+\beta_{1}
    \left(\varLambda;\omega\right)
    \frac{\partial}
    {\partial\omega}+
    \beta_{2}
    \left(\varLambda;\varkappa\right)
    \frac{\partial}
    {\partial\varkappa}\right\}
    \widetilde{\mathcal{S}}
    \left(\omega;\varkappa\right)=
    u\left(\varLambda;\omega;\varkappa
    \right)+t\left(\varLambda;\omega;
    \varkappa\right)
    \widetilde{\mathcal{S}}^{\,2}
    \!\left(\omega;\varkappa\right).
    \label{almost_self_similar_equation}
\end{equation}

In expression (\ref{almost_self_similar_equation}) we use the following compact notation for ``beta functions'' \cite{zinn1989field,vasil2004field}:
\begin{equation}
    \beta_{1}
    \left(\varLambda;\omega\right)=
    \varLambda\frac{\partial\omega
    \left(\varLambda;\omega\right)}
    {\partial\varLambda},\quad
    \beta_{2}
    \left(\varLambda;\varkappa\right)=
    \varLambda\frac{\partial\varkappa
    \left(\varLambda;\varkappa\right)}
    {\partial\varLambda}.
    \label{beta_functions_denotations}
\end{equation}

If beta functions and the coefficient functions
$u\left(\varLambda;\omega;\varkappa\right)$ and $t\left(\varLambda;\omega;\varkappa\right)$
\begin{equation}
    u\left(\varLambda;\omega;\varkappa
    \right)=\frac{\mathcal{U}\left(\varLambda;
    w,k\right)}{\varLambda^{a-1}}, \quad 
    t\left(\varLambda;\omega;\varkappa\right)=
    \varLambda^{a+1}\mathcal{T}\left(\varLambda;
    w;k\right)
    \label{coefficient_functions_denotations}
\end{equation}
don't explicitly depend on $\varLambda$, the equation (\ref{almost_self_similar_equation}) indeed has a self-similar solution. Let us consider power dependence of the functions $\omega$ and $\varkappa$ ($b$ and $c$ are arbitrary parameters):
\begin{equation}
    \omega\left(\varLambda;w\right)=
    \varLambda^{b}w,\quad
    \varkappa\left(\varLambda;k\right)=
    \varLambda^{c}k.
    \label{choise_of_w-k}
\end{equation}

In this case, the expression for the beta functions $\beta_{1}\left(\varLambda;\omega\right)$ and $\beta_{2}\left(\varLambda;\varkappa\right)$ reads as follows: 
\begin{equation}
    \beta_{1}
    \left(\varLambda;\omega\right)=
    \beta_{1}\left(\omega\right)=b\omega,\quad 
    \beta_{2}
    \left(\varLambda;\varkappa\right)=
    \beta_{2}
    \left(\varkappa\right)=c\varkappa.
\end{equation}

Thus, we arrive at the self-similar Riccati equation:
\begin{equation}
    \left\{b\omega
    \frac{\partial}
    {\partial\omega}+
    c\varkappa
    \frac{\partial}
    {\partial\varkappa}\right\}
    \widetilde{\mathcal{S}}
    \left(\omega;\varkappa\right)=
    u\left(\omega;\varkappa\right)-
    a\widetilde{\mathcal{S}}
    \left(\omega;\varkappa\right)
    +t\left(\omega;\varkappa\right)
    \widetilde{\mathcal{S}}^{\,2}
    \!\left(\omega;\varkappa\right).
    \label{final_self_similar_equation}
\end{equation}

For a given coefficient functions
$u\left(\omega;\varkappa\right)$ and $t\left(\omega;\varkappa\right)$ the equation (\ref{final_self_similar_equation}) can be solved. After that, one can consider a more general calculation of critical exponents. If the exact Riccati equation has equation (\ref{final_self_similar_equation}) as a limiting case, one can consider a perturbation of equation (\ref{final_self_similar_equation}) solution of the form:
\begin{equation}
    \mathcal{S}\left(\varLambda;w;k\right)=
    \varLambda^{a}
    \widetilde{\mathcal{S}}
    \left(\omega\left(\varLambda;w\right)\!;
    \varkappa\left(\varLambda;k\right)\right)+
    \delta\mathcal{S}\left(\varLambda;w;k\right).
    \label{critical exponents}
\end{equation}

The exact Riccati equation should be linearized with respect to the second term in the right-hand side of the expression (\ref{critical exponents}). This linearization determines the value of the critical exponents in the system. This is the standard technique of FRG \cite{kopbarsch,wipf2012statistical,rosten2012fundamentals,igarashi2009realization}: The rescaled form of the FRG flow equations (the corresponding Cauchy problem) is most convenient to discuss fixed points of the RG.

\subsection{Integration Formula for Functionals}

In this subsection, we derive the integration (Newton--Leibniz or NL type) formula for functionals. Detailed discussion of NL formula derivation is interesting by the following reason: The FRG flow equations (Wilson--Polchinski or Wetterich--Morris) are derived in a similar way \cite{kopbarsch}. Indeed, the introduction of any $t$-deformed functional with the subsequent differentiation with respect to scale $t$, and transformation of the obtained result into an expression in terms of functional derivatives with respect to the argument of functional are the key moments in the derivation of NL formula as well as different FRG flow equations. One can say that NL formula is a particular case of FRG.

Let us carry out the derivation for some functional $\mathcal{F}\left[\varLambda\right]$. First step: We add an increment $\varDelta_{t}$, depending on parameter $t$, to the argument $\varLambda$ of functional $\mathcal{F}\left[\varLambda\right]$. Second step: We can rewrite the increment in terms of functional translation operator. Third step: By taking the derivative of the following expression with respect to scale $t$ and using the functional translation operator once again, we obtain the expression:
\begin{equation}
    \mathcal{F}
    \left[\varLambda+\varDelta_{t}\right]=
    e^{\left(\varDelta_{t}\left|\frac{\delta}
    {\delta\varLambda}\right.\right)}
    \mathcal{F}\left[\varLambda\right],\quad
    \frac{\partial\mathcal{F}
    \left[\varLambda+\varDelta_{t}\right]}
    {\partial t}=
    \left(\frac{\partial\varDelta_{t}}{\partial t}
    \bigg|\frac{\delta\mathcal{F}
    \left[\varLambda+\varDelta_{t}\right]}
    {\delta\varLambda}\right).
    \label{ch3-eq12}
\end{equation}

Let us integrate the equation (\ref{ch3-eq12}) over $t$ in the limits from $t_{0}$ to $t_{1}$. The result of integration reads:
\begin{equation}
    \mathcal{F}\left[\varLambda+\varDelta_{t_{1}}\right]-
    \mathcal{F}\left[\varLambda+\varDelta_{t_{0}}\right]=
    \int\limits_{t_{0}}^{t_{1}}dt
    \int d^{D}z\,\frac{\partial\varDelta_{t}
    \left(z\right)}{\partial t}
    \frac{\delta\mathcal{F}
    \left[\varLambda+\varDelta_{t}\right]}
    {\delta\varLambda\left(z\right)}.
    \label{ch3-eq13}
\end{equation}

To obtain the NL formula in a canonical form, the value $\varDelta_{t}$ should be chosen in the simplest form which is the product of $t$ and $\varDelta$, i.e. $\varDelta_{t}=t\varDelta$. The dimensions of $\varDelta_{t}$ and $\varDelta$ are coincide because the parameter $t$ is dimensionless. In this case, we obtain the following expression:
\begin{equation}
     \mathcal{F}\left[\varLambda+t_{1}\varDelta\right]-
    \mathcal{F}\left[\varLambda+t_{0}\varDelta\right]=
    \int\limits_{t_{0}}^{t_{1}}dt
    \int d^{D}z\,\varDelta\left(z\right)
    \frac{\delta\mathcal{F}
    \left[\varLambda+\varDelta_{t}\right]}
    {\delta\varLambda\left(z\right)}.
    \label{ch3-eq14}
\end{equation}

Let us choose $t_{0}=0$ and $t_{1}=1$. In this case we arrive to the canonical NL formula for functionals:
\begin{equation}
    \mathcal{F}\left[\varLambda+\varDelta\right]-
    \mathcal{F}\left[\varLambda\right]=
    \int\limits_{0}^{1}dt
    \int d^{D}z\,\varDelta\left(z\right)
    \frac{\delta\mathcal{F}
    \left[\varLambda+t\varDelta\right]}
    {\delta\varLambda\left(z\right)}.
    \label{ch3-eq15}
\end{equation}

The obtained integration formula (\ref{ch3-eq15}) reconstruct the functional by its first functional derivative. For the self-consistency of obtained construction, several additional conditions have to be satisfied. In particular, the second functional derivative of the functional $\mathcal{F}\left[\varLambda\right]$ has to be independent of the variation order. These conditions are satisfied in all the constructions considered in the present paper.

The practical application of the integration formula (\ref{ch3-eq15}) is based on two additional expressions: The shift of variables and the functional equation in terms of the first variational derivative:
\begin{equation}
    \frac{\delta\mathcal{F}
    \left[\varLambda+t\Delta\right]}
    {\delta\left(\varLambda+t\Delta\right)
    \left(z\right)}=
    \frac{\delta\mathcal{F}
    \left[\varLambda+t\Delta\right]}
    {\delta\varLambda
    \left(z\right)},\quad
    \frac{\delta\mathcal{F}
    \left[\varLambda\right]}
    {\delta\varLambda\left(z\right)}=
    \mathcal{G}
    \left[\varLambda\right]\left(z\right).
    \label{G-introduction}
\end{equation}

Substituting expression (\ref{G-introduction}) into (\ref{ch3-eq15}) we obtain the formula for the reconstruction of the functional from its first variational derivative:
\begin{equation}
    \mathcal{F}
    \left[\varLambda+\varDelta\right]-
    \mathcal{F}\left[\varLambda\right]=
    \int\limits_{0}^{1}dt
    \int d^{D}z\,
    \varDelta\left(z\right)
    \mathcal{G}\left[\varLambda+t\varDelta
    \right]\left(z\right).
    \label{The_reconstruction_formula1}
\end{equation}

In the following subsections, we will repeatedly use the reconstruction formula (\ref{The_reconstruction_formula1}) to prove an important result: The solutions and conclusions obtained for the delta-field configuration of $\varLambda$ remain valid for a wide class of field configurations.

\subsection{Translation-Invariant Functional Solution for Green Function}

Let us use the reconstruction formula (\ref{The_reconstruction_formula1}) to integrate functional equation (\ref{Equation_Two-Particle_GF_TIP_MR}):
\begin{equation}
    \mathcal{S}\left[\varLambda+
    \varDelta\right]\left(k\right)-
    \mathcal{S}\left[\varLambda\right]
    \left(k\right)=\int\limits_{0}^{1}dt
    \left\{\mathcal{U}\left[\varLambda;
    \varDelta\right]\left(t;k\right)+
    \mathcal{T}\left[\varLambda;
    \varDelta\right]\left(t;k\right)
    |\mathcal{S}\left[\varLambda+t\varDelta
    \right]\left(k\right)|^{2}\right\}.
    \label{new_translation_invariant_solution1}
\end{equation} 

The functionals in the right-hand side of the equation (\ref{The_reconstruction_formula1}) are defined as follows:
\begin{equation}
\begin{split}
    &\mathcal{U}\left[\varLambda;
    \varDelta\right]\left(t;k\right)
    \equiv\int d^{D}z\,\varDelta\left(z\right)
    \mathcal{U}\left[\varLambda+
    t\varDelta\right]\left(z;k\right),\\
    &\mathcal{T}\left[\varLambda;
    \varDelta\right]\left(t;k\right)
    \equiv\int d^{D}z\,\varDelta\left(z\right)
    \mathcal{T}\left[\varLambda+
    t\varDelta\right]\left(z;k\right).
    \label{new_translation_invariant_solution2}
\end{split}
\end{equation}

The functionals (\ref{new_translation_invariant_solution2}) are the functional coefficients of the equation (\ref{new_translation_invariant_solution1}). Also let us make one remark: The functions $\varLambda$ and $\varDelta$ depend on the spatial coordinate $z$. The functionals $\mathcal{U}$ and $\mathcal{T}$ on the right-hand side of the equalities (\ref{new_translation_invariant_solution2}) can explicitly depend on the same coordinate $z$. In what follows, we will choose delta-field configurations for the functions $\varLambda$ and $\varDelta$. The corresponding amplitudes we also denote as $\varLambda$ and $\varDelta$. The expression (\ref{For_The_Reviewer_2}) will clarify this in more detail.

Now, in order to develop our intuition regarding the functionals (\ref{new_translation_invariant_solution2}), we obtain the limiting solution for $\mathcal{S}$ when $\|\varLambda\|\ \rightarrow +\infty $, $\|\varDelta\| < +\infty$. This limiting case is important because the functional Taylor series is an expansion in powers of the field $\varLambda$. The other side asymptotics is of special interest, and this asymptotics reads:
\begin{equation}
    |\mathcal{S}\left[\varLambda\right]
    \left(k\right)|=\sqrt{-\frac{\int
    d^{D}z\,\varDelta\left(z\right)
    \mathcal{U}\left[\varLambda\right]
    \left(z;k\right)}{\int d^{D}z\,\varDelta
    \left(z\right)\mathcal{T}
    \left[\varLambda\right]\left(z;k\right)}}.
    \label{new_translation_invariant_solution3}
\end{equation}

Further, using The Mean Value Theorem with $\tau \in (0,1)$, one can represent the equation (\ref{new_translation_invariant_solution1}) in the following form:
\begin{equation}
    \mathcal{S}\left[\varLambda+
    \varDelta\right]\left(k\right)-
    \mathcal{S}\left[\varLambda\right]
    \left(k\right)=
    \int\limits_{0}^{1}dt\,
    \mathcal{U}\left[\varLambda;
    \varDelta\right]\left(t;k\right)+
    |\mathcal{S}
    \left[\varLambda+\tau\varDelta
    \right]\left(k\right)|^{2}
    \int\limits_{0}^{1}dt\,
    \mathcal{T}\left[\varLambda;
    \varDelta\right]\left(t;k\right).
    \label{Mean_value_theorem}
\end{equation}

Expression (\ref{Mean_value_theorem}) is a source of 
various estimates for the functional equation (\ref{new_translation_invariant_solution1}). It is also a starting point for an approximate solution of the equation (\ref{new_translation_invariant_solution1}). 

Next, consider delta-field configurations of $\varLambda$ and $\varDelta$ (the corresponding amplitudes we also denote as $\varLambda$ and $\varDelta$). Thus, we choose certain directions in the space of functions:
\begin{equation}
    \varLambda\left(z\right)=\varLambda\,
    \delta^{(D)}\left(z-w\right)
    \equiv\varLambda_{\delta,w},\quad \varDelta\left(z\right)=\varDelta\,
    \delta^{(D)}\left(z-w\right)
    \equiv\varDelta_{\delta,w}.
    \label{For_The_Reviewer_2}
\end{equation}

Thus, additionally setting $\varLambda=\varLambda_{0}$, $\varDelta=\varLambda_{1}-\varLambda_{0}$ and $t\varDelta=\varLambda_{t}-\varLambda_{0}$, the functional equation (\ref{new_translation_invariant_solution1}) becomes an integral equation for some function $\mathcal{S}\left(\varLambda_{t};w;k\right)$:
\begin{equation}
    \mathcal{S}\left(\varLambda_{1};w;k\right)-
    \mathcal{S}\left(\varLambda_{0};w;k\right)=
    \int\limits_{\varLambda_{0}}^{\varLambda_{1}}
    d\varLambda_{t}
    \left\{\mathcal{U}
    \left(\varLambda_{t};w;k\right)+
    \mathcal{T}
    \left(\varLambda_{t};w;k\right)
    |\mathcal{S}
    \left(\varLambda_{t};w;k\right)|^{2}\right\}.
    \label{new_translation_invariant_solution4}
\end{equation}

The expression for the function $\mathcal{S}\left(\varLambda_{t};w;k\right)$ appearing in the integral equation (\ref{new_translation_invariant_solution4}) reads:
\begin{equation}
    \mathcal{S}\left(\varLambda_{t};w;k\right)=
    \mathcal{S}^{(0)}\left(k\right)+
    \sum_{n=1}^{\infty}\frac{\varLambda_{t}^{n}}{n!}\,
    \mathcal{S}^{(n)}\left(w,\ldots,w;k\right),\quad \varLambda_{t}=
    t\left(\varLambda_{1}-\varLambda_{0}\right)+
    \varLambda_{0}.
    \label{new_translation_invariant_solution5}
\end{equation}

We now differentiate the integral equation (\ref{new_translation_invariant_solution4}) with respect to the upper limit of integral $\varLambda_{1}$, assuming the lower limit $\varLambda_{0}$ to be constant. As a result, we come to the Riccati equations (\ref{Equation_Two-Particle_GF_TIP_Delta}) or (\ref{Riccati_Equation_Two-Particle_GF_TIP}), that were obtained earlier in a different way. Thus, the derivation above is an independent verification of previously obtained results. In the next subsection, we get an equation for the opposite field configurations of $\varLambda$ and $\varDelta$ -- constant fields. This allows us to make an important conclusion about the nature of the functional equation (\ref{new_translation_invariant_solution1}).

\subsection{Translation-Invariant Solution for Green Function on Constant-Field Configuration}

Traveling back to the functional equation (\ref{new_translation_invariant_solution1}) and its functional coefficients (\ref{new_translation_invariant_solution2}), we consider constant-field configurations: $\varLambda\left(z\right)=\varLambda$ and $\varDelta\left(z\right)=\varDelta$. As before, set $\varLambda=\varLambda_{0}$, $\varDelta=\varLambda_{1}-\varLambda_{0}$ and $t\varDelta=\varLambda_{t}-\varLambda_{0}$. The functional equation (\ref{new_translation_invariant_solution1}) becomes an integral equation for some function $\mathcal{S}\left(\varLambda_{t};k\right)$:
\begin{equation}
    \mathcal{S}
    \left(\varLambda_{1};k\right)-
    \mathcal{S}
    \left(\varLambda_{0};k\right)=
    \int\limits_{\varLambda_{0}}^{\varLambda_{1}}
    d\varLambda_{t}
    \int d^{D}z
    \left\{\mathcal{U}
    \left(\varLambda_{t};z;k\right)+
    \mathcal{T}
    \left(\varLambda_{t};z;k\right)
    |\mathcal{S}
    \left(\varLambda_{t};k\right)|^{2}\right\}.
    \label{new_translation_invariant_solution6}
\end{equation} 

The expression for the function $\mathcal{S}\left(\varLambda_{t};k\right)$ appearing in the integral equation (\ref{new_translation_invariant_solution6}) reads:
\begin{equation}
    \mathcal{S}
    \left(\varLambda_{t};k\right)=
    \mathcal{S}^{(0)}\left(k\right)+
    \sum_{n=1}^{\infty}
    \frac{\varLambda_{t}^{n}}{n!}
    \int d^D z_{1}\ldots\int d^D z_{n}\, \mathcal{S}^{(n)}
    \left(z_{1},\ldots,z_{n};k\right).
    \label{new_translation_invariant_solution7}
\end{equation}

We now differentiate the integral equation (\ref{new_translation_invariant_solution6}) with respect to the upper limit of integral $\varLambda_{1}$, assuming the lower limit $\varLambda_{0}$ to be constant. As a result, we come to the following equation:
\begin{equation}
    \frac{\partial\mathcal{S}
    \left(\varLambda_{1};k\right)}
    {\partial\varLambda_{1}}=
    \int d^{D}z\,\mathcal{U}
    \left(\varLambda_{1};z;k\right)+
    |\mathcal{S}
    \left(\varLambda_{1};k\right)|^{2}
    \int d^{D}z\,\mathcal{T}
    \left(\varLambda_{1};z;k\right).
    \label{new_translation_invariant_solution8}
\end{equation}

If coefficient functions $\mathcal{U}\left(\varLambda_{1};z;k\right)$ and $\mathcal{T}\left(\varLambda_{1};z;k\right)$ are even functions of $k$, we obtain closed with respect to the function $\mathcal{S}\left(\varLambda_{1};k\right)$ Riccati equation:
\begin{equation}
    \frac{\partial\mathcal{S}
    \left(\varLambda_{1};k\right)}
    {\partial\varLambda_{1}}=
    \mathcal{U}
    \left(\varLambda_{1};k\right)+
    \mathcal{T}
    \left(\varLambda_{1};k\right)
    \mathcal{S}^{2}
    \left(\varLambda_{1};k\right).
    \label{new_translation_invariant_solution9}
\end{equation}

The functions in the right-hand side of the equation (\ref{new_translation_invariant_solution9}) are defined as follows:
\begin{equation}
    \mathcal{U}
    \left(\varLambda_{1};k\right)\equiv
    \int d^{D}z\,\mathcal{U}
    \left(\varLambda_{1};z;k\right),\quad
    \mathcal{T}
    \left(\varLambda_{1};k\right)\equiv
    \int d^{D}z\,\mathcal{T}
    \left(\varLambda_{1};z;k\right).
    \label{new_translation_invariant_solution10}
\end{equation}

We got the Riccati equation again. Thus, in two limiting cases (delta-field and constant-field configurations of $\varLambda$ and $\varDelta$), we obtain the Riccati equation for the two-particle holographic GF. This doesn't prove that the Riccati equation can be obtained for any configuration of $\varLambda$ and $\varDelta$ from the corresponding space of fields. However, according to the results obtained, we can assert that such a space indeed exists. Thus, the equation (\ref{new_translation_invariant_solution1}) is ``stable'' with respect to the choice of fields $\varLambda$ and $\varDelta$. At the end of this subsection, we make one remark. Technically, equations  (\ref{new_translation_invariant_solution8}) and (\ref{new_translation_invariant_solution9}) can be obtained by integrating the functional equation (\ref{Equation_Two-Particle_GF_TIP_MR}) over $z$, and then setting constant-field configuration of $\varLambda$.

\subsection{Separable Solution for Green Function on Delta-Field Configuration}

At the end of the Functional Hamilton--Jacobi Equation and its Hierarchy section let us consider the separable problem. The separable problem is often encountered in QM models of Nuclear Physics and in Theory of Superconductivity. In the coordinate representation this problem reads as follows:
\begin{equation}
    \mathcal{T}^{(0)}\left[\varLambda\right]
    \left(z;y_{1},y_{2}\right)=
    \mathcal{T}\left[\varLambda\right]
    \left(z;y_{1}\right)
    \mathcal{T}\left[\varLambda\right]
    \left(z;y_{2}\right),\quad
    \mathcal{U}^{(2)}\left[\varLambda\right]
    \left(z;y_{1},y_{2}\right)=0.
    \label{Separable_problem}
\end{equation}

Since the symmetry of the solution often preserves the symmetry of the equation coefficients, we consider separable two-particle holographic GF in the coordinate representation:
\begin{equation}
    \mathcal{S}^{(2)}\left[\varLambda\right]
    \left(y_{1},y_{2}\right)=
    \mathcal{S}\left[\varLambda\right]
    \left(y_{1}\right)
    \mathcal{S}\left[\varLambda\right]
    \left(y_{2}\right).
    \label{Separable_solution}
\end{equation}

Further, the equation (\ref{Equation_Two-Particle_GF_CR}) for two-particle holographic GF for separable problem (\ref{Separable_problem}) in coordinate representation is:
\begin{equation}
    \frac{\delta}{\delta\varLambda\left(z\right)}
    \left\{\mathcal{S}\left[\varLambda\right]
    \left(y_{1}\right)
    \mathcal{S}\left[\varLambda\right]
    \left(y_{2}\right)\right\}=
    \mathcal{S}\left[\varLambda\right]
    \left(y_{1}\right)
    \mathcal{S}\left[\varLambda\right]
    \left(y_{2}\right)
    \left\{\int d^D x
    \mathcal{T}\left[\varLambda\right]
    \left(z;x\right)
    \mathcal{S}\left[\varLambda\right]
    \left(x\right)\right\}^{2}.
    \label{Equation_Two-Particle_GF_SP_CR_1}
\end{equation}

Traveling back, the equation (\ref{Equation_Two-Particle_GF_SP_CR_1}) for two-particle holographic GF can be converted as follows:
\begin{equation}
    \frac{\delta\mathcal{S}^{(2)}
    \left[\varLambda\right]
    \left(y_{1},y_{2}\right)}
    {\delta\varLambda\left(z\right)}=
    \mathcal{K}^{2}
    \left[\varLambda\right]\left(z\right)
    \mathcal{S}^{(2)}
    \left[\varLambda\right]
    \left(y_{1},y_{2}\right).
    \label{Equation_Two-Particle_GF_SP_CR_2}
\end{equation}

The expression for the functional $\mathcal{K}$ appearing in the equation (\ref{Equation_Two-Particle_GF_SP_CR_2}) reads:
\begin{equation}
    \mathcal{K}\left[\varLambda\right]
    \left(z\right)=\int d^D x
    \mathcal{T}\left[\varLambda\right]
    \left(z;x\right)
    \mathcal{S}\left[\varLambda\right]
    \left(x\right).
    \label{Def_K_functional}
\end{equation}

Taking into account the results obtained in the previous subsections, for further derivation of the equation (\ref{Equation_Two-Particle_GF_SP_CR_2}) solution we choose a strategy based on the delta-field configuration of $\varLambda$. For this reason, consider two-particle holographic GF $\mathcal{S}$ as a functional of the field $\varLambda$ and its functional Taylor series over $\varLambda$ in the coordinate representation:
\begin{equation}
    \mathcal{S}\left[\varLambda\right]\left(y\right)=
    \mathcal{S}^{(0)}\left(y\right)+
    \sum_{n=1}^{\infty}\frac{1}{n!}
    \int d^D z_{1}\ldots\int d^D z_{n}\, \mathcal{S}^{(n)}\left(z_{1},\ldots,z_{n};y\right)
    \varLambda\left(z_{1}\right)\ldots
    \varLambda\left(z_{n}\right).
    \label{FTS_Two-Particle_GF_CR_Separable}
\end{equation}

Variational derivative of expression (\ref{FTS_Two-Particle_GF_CR_Separable}) reads as follows:
\begin{equation}
    \frac{\delta\mathcal{S}
    \left[\varLambda\right]\left(y\right)}
    {\delta\varLambda\left(z\right)}=
    \mathcal{S}^{(1)}\left(z;y\right)+
    \sum_{n=2}^{\infty}\frac{1}{\left(n-1\right)!}
    \int d^D z_{2}\ldots\int d^D z_{n}\, \mathcal{S}^{(n)}\left(z,z_{2},\ldots,z_{n};y\right)
    \varLambda\left(z_{2}\right)\ldots
    \varLambda\left(z_{n}\right).
    \label{Der_FTS_Two-Particle_GF_CR_Separable}
\end{equation}

For the delta-field configuration of $\varLambda$ the expression (\ref{FTS_Two-Particle_GF_CR_Separable}) is:
\begin{equation}
    \mathcal{S}\left[\varLambda\right]\left(y\right)
    \big\rvert_{\varLambda_{\delta,w}}=
    \mathcal{S}^{(0)}\left(y\right)+
    \sum_{n=1}^{\infty}\frac{\varLambda^{n}}{n!}\,
    \mathcal{S}^{(n)}\left(w,\ldots,w;y\right)\equiv \mathcal{S}\left(\varLambda;w;y\right).
    \label{FTS_Two-Particle_GF_Delta_Separable}
\end{equation}

For the delta-field configuration of $\varLambda$ the expression (\ref{Der_FTS_Two-Particle_GF_CR_Separable}) reads:
\begin{equation}
    \frac{\delta\mathcal{S}
    \left[\varLambda\right]\left(y\right)}
    {\delta\varLambda\left(z\right)}
    \bigg\rvert_{\varLambda_{\delta,w}}=
    \mathcal{S}^{(1)}\left(z;y\right)+
    \sum_{n=2}^{\infty}\frac{\varLambda^{n-1}}
    {\left(n-1\right)!}\, \mathcal{S}^{(n)}\left(z,w,\ldots,w;y\right).
    \label{Der_FTS_Two-Particle_GF_Delta_Separable}
\end{equation}

The expressions (\ref{FTS_Two-Particle_GF_Delta_Separable})--(\ref{Der_FTS_Two-Particle_GF_Delta_Separable}) demonstrate the following fact: If we choose delta-field configuration of $\varLambda$ in the equation (\ref{Equation_Two-Particle_GF_SP_CR_2}) and then set $z=w$, the equation (\ref{Equation_Two-Particle_GF_SP_CR_2}) becomes closed (up to $k\rightarrow -k$ symmetry) with respect to the product of functions $\mathcal{S}\left(\varLambda;w;y\right)$ (\ref{FTS_Two-Particle_GF_Delta_Separable}):
\begin{equation}
    \frac{\partial\mathcal{S}^{(2)}
    \left(\varLambda;w;y_{1},y_{2}\right)}
    {\partial\varLambda}=
    \mathcal{K}^{2}
    \left(\varLambda;w\right)
    \mathcal{S}^{(2)}
    \left(\varLambda;w;y_{1},y_{2}\right).
    \label{Equation_Two-Particle_GF_SP_Delta}
\end{equation}

The equation (\ref{Equation_Two-Particle_GF_SP_Delta}) is a pseudo-linear equation: The coefficient $\mathcal{K}$ of the equation (\ref{Equation_Two-Particle_GF_SP_Delta}) depends on the function $\mathcal{S}^{(2)}$, what follows from the equality (\ref{Def_K_functional}). Such an equation can be rewritten in integral form, repeating the solution to the linear equation ($\mathcal{C}$ is a ``constant'' of integration -- function that doesn't depend on $\varLambda$):
\begin{equation}
    \mathcal{S}^{(2)}
    \left(\varLambda;w;y_{1},y_{2}\right)=
    \mathcal{C}\left(w;y_{1},y_{2}\right)
    e^{\int_{\varLambda_{0}}^{\varLambda}d\varLambda'
    \mathcal{K}^{2}\left(\varLambda',w\right)}.
    \label{Two-Particle_GF_SP_Delta_Solution}
\end{equation}

The function $\mathcal{C}$ must be separable:
\begin{equation}
    \mathcal{C}\left(w;y_{1},y_{2}\right)=
    \mathcal{C}\left(w;y_{1}\right)
    \mathcal{C}\left(w;y_{2}\right).
    \label{C_Function}
\end{equation}

As soon as, according to the definition (\ref{Def_K_functional}),
\begin{equation}
    \mathcal{K}\left(\varLambda;w\right)=
    \int d^D x
    \mathcal{T}\left(\varLambda;w;x\right)
    \mathcal{S}\left(\varLambda;w;x\right),
    \label{K_functional_Delta}
\end{equation}
applying integral operator with kernel
\begin{equation}
    \int d^D y_{1}\int d^D y_{2}
    \mathcal{T}\left(\varLambda;w;y_{1}\right)
    \mathcal{T}\left(\varLambda;w;y_{2}\right)
    \label{Operator_Separable}
\end{equation}
to both sides of the integral equation (\ref{Two-Particle_GF_SP_Delta_Solution}), we obtain a closed integral equation for the function $\mathcal{K}^{2}$:
\begin{equation}
    \mathcal{K}^{2}\left(\varLambda;w\right)=
    \mathcal{C}^{2}_{1}\left(\varLambda;w\right)
    e^{\int_{\varLambda_{0}}^{\varLambda}d\varLambda'
    \mathcal{K}^{2}\left(\varLambda',w\right)}.
    \label{K_SP_Delta_Pseudo-Solution_1}
\end{equation}

The expression for the function $\mathcal{C}_{1}$ appearing in the equation (\ref{K_SP_Delta_Pseudo-Solution_1}) reads:
\begin{equation}
    \mathcal{C}_{1}\left(\varLambda;w\right)=
    \int d^D y\mathcal{T}\left(\varLambda;w;y\right)
    \mathcal{C}\left(w;y\right).
    \label{C_Int_Function}
\end{equation}

The logarithmic derivative of expression (\ref{K_SP_Delta_Pseudo-Solution_1})
\begin{equation}
    \frac{\partial
    \ln{\mathcal{K}^{2}\left(\varLambda;w\right)}}
    {\partial\varLambda}=
    \frac{\partial
    \ln{\mathcal{C}^{2}_{1}\left(\varLambda;w\right)}}
    {\partial\varLambda}+
    \mathcal{K}^{2}\left(\varLambda;w\right).
    \label{K_SP_Delta_Pseudo-Solution_2}
\end{equation}
we arrive at the Bernoulli differential equation. The latter is reduced to the linear differential equation by replacing functions and therefore can be integrated by quadratures. The solution of the equation (\ref{K_SP_Delta_Pseudo-Solution_2}) reads as follows:
\begin{equation}
    \mathcal{K}^{2}\left(\varLambda;w\right)=
    \frac{\mathcal{C}^{2}_{1}\left(\varLambda;w\right)}
    {\mathcal{C}_{2}\left(w\right)
    \mathcal{C}^{2}_{1}\left(\varLambda_{0};w\right)
    -\int_{\varLambda_{0}}^{\varLambda}d\varLambda'
    \mathcal{C}^{2}_{1}\left(\varLambda';w\right)},\quad
    \mathcal{C}_{2}\left(w\right)>0.
    \label{For_The_Reviewer_2_2}
\end{equation}

New ``constant'' of integration $\mathcal{C}_{2}$ -- function that depends on $w$. Traveling back, the solution for the two-particle holographic GF $\mathcal{S}^{(2)}$ is obtained (the expression (\ref{For_The_Reviewer_2_2}) gives the function $\mathcal{K}$, the expression (\ref{C_Int_Function}) gives the function $\mathcal{C}_{1}$, finally the expression (\ref{Two-Particle_GF_SP_Delta_Solution}) gives the function $\mathcal{S}^{(2)}$). In conclusion of the section, let us note that the functional equation (\ref{Equation_Two-Particle_GF_SP_CR_2}) can also be investigated using the reconstruction formula for functionals (\ref{The_reconstruction_formula1}). We have established that the separable problem (\ref{Separable_problem}) has the property of ``stability'' with respect to the choice of fields $\varLambda$ and $\varDelta$ as well as the translation-invariant problem (\ref{Translation-invariant_problem}).

\section{Functional Schr\"{o}dinger Equation and Semiclassical Approximation}

In this section we consider the functional Schr\"{o}dinger equation. As in the part devoted to the functional HJ equation we widely use condensed notation for clarity \cite{kopbarsch}. In this section we present a rigorous derivation of the quantum functional HJ equation, containing the quantum correction term, and continuity equation from the functional Schr\"{o}dinger equation and provide a proof of the functional semiclassical approximation (semiclassics).

\subsection{Functional Schr\"{o}dinger Equation}

First of all, let us make an important remark about the difference between QM and QFT. In QM, the wave function $\varPsi$ is an ``ordinary'' function of several variables. These variables are the spatial coordinate $x$ and the time $t$. In QFT, the role of the wave function is played by the wave functional $\varPsi$. This functional  depends on its arguments -- the fields $\varLambda$ (analogue of time $t$) and $\varphi$ (analogue of spatial coordinate $x$). The spatial coordinate $x$ and the time $t$ are now the arguments of the fields $\varLambda$ and $\varphi$ (in this paper, we consider the time as one of the spatial coordinates in $D$-dimensional Euclidean space). For this reason the Schr\"{o}dinger equation in QM is a partial differential equation, while the Schr\"{o}dinger equation in QFT is an equation in variational derivatives. This functional Schr\"{o}dinger equation in the coordinate representation is given by the following expression:
\begin{equation}
    \frac{\delta\varPsi
    \left[\varLambda,\varphi\right]}
    {\delta\varLambda\left(z\right)}=
    i\hat{\mathcal{H}}
    \left(z\right)\varPsi
    \left[\varLambda,\varphi\right],\quad
    \hat{\mathcal{H}}\left(z\right)=
    \mathcal{H}
    \left[\varLambda,\varphi,\hat{\varpi}\right]
    \left(z\right),\quad \hat{\varpi}\left(x\right)=\frac{\delta}
    {\delta\varphi\left(x\right)}.
    \label{FS_Equation_CR}
\end{equation}

The functional $\varPsi$ appearing in the expression (\ref{FS_Equation_CR}) is the wave functional, $\varphi$ is the scalar field which is the function of the spatial coordinate $x$, $\hat{\varpi}$ up to unit imaginary number is the functional momentum operator. In a general case, the functional Hamiltonian $\hat{\mathcal{H}}$ can be presented as the sum of the functional Taylor series, but we are going to consider a more specific case of the sum of kinetic and potential energy functionals as in standard QM:
\begin{equation}
    \hat{\mathcal{H}}\left(z\right)=
    \int d^{D}x_{1}\int d^{D}x_{2}\,
    \mathcal{K}\left[\varLambda, \varphi\right]\left(z;x_{1},x_{2}\right)
    \hat{\varpi}\left(x_{1}\right)
    \hat{\varpi}\left(x_{2}\right)+
    \mathcal{U}
    \left[\varLambda,\varphi\right]
    \left(z\right).
    \label{FS_Hamiltonian_CR}
\end{equation}

As in the case of the functional HJ equation, all the spatial variables $x,y,z\in\mathbb{R}^{D}$ although, from geometrical point of view, $z$ can refer to a different space with different $D$. The functional $\mathcal{K}$ is a generalization of the inverse mass tensor (with a minus sign, which is convenient in this paper). The inverse mass tensor itself arises in problems of classical and quantum mechanics. For example, the Schr\"{o}dinger equation with the inverse mass tensor is often occurs in problems of the quantum theory of condensed matter. Also, in problems of classical and quantum mechanics, this tensor can depend on coordinates (in the case of functional generalization, on fields). For example, the Schr\"{o}dinger equation with $x$-dependent mass also occurs in problems of the quantum theory of condensed matter.

The potential energy $\mathcal{U}$ in classical and quantum mechanics is an arbitrary function of spatial coordinate $x$ and, possibly, time $t$. This function isn't convoluted with any degree of momentum (in contrast to the tensor of inverse masses). A typical example of this function is a fourth degree polynomial -- anharmonic oscillator. In the case of the functional generalization, the potential energy functional $\mathcal{U}$ depends on the fields $\varLambda$ (analogue of time $t$) and $\varphi$ (analogue of spatial coordinate $x$). No integration over $x_{1}$ and $x_{2}$ in the term corresponding to the potential energy $\mathcal{U}$ in the expression (\ref{FS_Hamiltonian_CR}) is required. If we choose the functional $\mathcal{U}$ in the form of the fourth degree polynomial (with respect to the field $\varphi$), we obtain the functional generalization of the anharmonic oscillator.

In the condensed notation expressions (\ref{FS_Equation_CR})--(\ref{FS_Hamiltonian_CR}) are more transparent (recall that the functional Hamiltonian $\hat{\mathcal{H}}$ depends on all values of the functional momentum operator $\hat{\varpi}$, therefore, the argument of $\hat{\varpi}$ is not specified in $\hat{\mathcal{H}}$):
\begin{equation}
    \frac{\delta\varPsi
    \left[\varLambda,\varphi\right]}
    {\delta\varLambda_{\alpha}}=
    i\hat{\mathcal{H}}_{\alpha}\varPsi
    \left[\varLambda,\varphi\right],\quad
    \hat{\mathcal{H}}_{\alpha}=
    \mathcal{H}_{\alpha}
    \left[\varLambda,\varphi,\hat{\varpi}\right],\quad \hat{\varpi}_{\mu}=\frac{\delta}
    {\delta\varphi_{\mu}},
    \label{FS_Equation_CN}
\end{equation}
where the functional Hamiltonian $\hat{\mathcal{H}}$
\begin{equation}
    \hat{\mathcal{H}}_{\alpha}=
    \int_{\mu_{1}}\int_{\mu_{2}}
    \mathcal{K}_{\alpha,\mu_{1}\mu_{2}}
    \left[\varLambda,\varphi\right]
    \hat{\varpi}_{\mu_{1}}\hat{\varpi}_{\mu_{2}}+
    \mathcal{U}_{\alpha}
    \left[\varLambda,\varphi\right].
    \label{FS_Hamiltonian_CN}
\end{equation}

\subsection{Derivation of Quantum Hamilton--Jacobi and Continuity Functional Equations and Semiclassics}

Assuming $\varPsi\left[\varLambda,\varphi\right]\in\mathbb{C}$, let $\varPsi\left[\varLambda,\varphi\right]=
\mathcal{A}\left[\varLambda,\varphi\right]
e^{i\mathcal{S}\left[\varLambda, \varphi\right]}$, where $\mathcal{A},\mathcal{S}\in\mathbb{R}$ and $\mathcal{A}>0$. So, variational derivatives of the wave functional in the condensed notation are (dots mean higher-order variational derivatives):
\begin{equation}
\begin{split}
    \frac{\delta\varPsi
    \left[\varLambda,\varphi\right]}
    {\delta\varLambda_{\alpha}}&=
    e^{i\mathcal{S}\left[\varLambda,\varphi\right]}
    \left\{\frac{\delta\mathcal{A}
    \left[\varLambda,\varphi\right]}
    {\delta\varLambda_{\alpha}}+
    i\mathcal{A}\left[\varLambda,\varphi\right]
    \frac{\delta\mathcal{S}
    \left[\varLambda,\varphi\right]}
    {\delta\varLambda_{\alpha}}\right\},\\
    \frac{\delta\varPsi
    \left[\varLambda,\varphi\right]}
    {\delta\varphi_{\mu}}&=
    e^{i\mathcal{S}\left[\varLambda,\varphi\right]}
    \left\{\frac{\delta\mathcal{A}
    \left[\varLambda,\varphi\right]}
    {\delta\varphi_{\mu}}+
    i\mathcal{A}\left[\varLambda,\varphi\right]
    \frac{\delta\mathcal{S}
    \left[\varLambda,\varphi\right]}
    {\delta\varphi_{\mu}}\right\}\ldots
\end{split}
\end{equation}

Omitting in notation $\left[\varLambda,\varphi\right]$ sign for explicit functional dependence on $\varLambda$ and $\varphi$ further in this subsection, factoring $e^{i\mathcal{S}}$ outside the integral sign (since $\mathcal{S}$ doesn't depend on either $\mu_{1}$ or $\mu_{2}$) in (\ref{FS_Equation_CN})--(\ref{FS_Hamiltonian_CN}) and dividing by it yields (from now on we consider $\varphi$-independent functional $\mathcal{K}$ for simplicity of further calculations):
\begin{equation}
\begin{split}
    \frac{\delta\mathcal{A}}
    {\delta\varLambda_{\alpha}}+i\mathcal{A}\,
    \frac{\delta\mathcal{S}}
    {\delta\varLambda_{\alpha}}&=
    \int_{\mu_{1}}\int_{\mu_{2}}
    \mathcal{K}_{\alpha,\mu_{1}\mu_{2}}
    \bigg\{-\mathcal{A}\,
    \frac{\delta^{2}\mathcal{S}}
    {\delta\varphi_{\mu_{1}}
    \delta\varphi_{\mu_{2}}}-
    2\frac{\delta\mathcal{A}}
    {\delta\varphi_{\mu_{1}}}
    \frac{\delta\mathcal{S}}
    {\delta\varphi_{\mu_{2}}}+\\ &+i\frac{\delta^{2}\mathcal{A}}
    {\delta\varphi_{\mu_{1}}
    \delta\varphi_{\mu_{2}}}-
    i\mathcal{A}\,
    \frac{\delta\mathcal{S}}
    {\delta\varphi_{\mu_{1}}}
    \frac{\delta\mathcal{S}}
    {\delta\varphi_{\mu_{2}}}\bigg\}+
    i\mathcal{A}\,\mathcal{U}_{\alpha}.
    \label{before_cont_and_HJ}
\end{split}
\end{equation}

The real part of the expression (\ref{before_cont_and_HJ}) yields the functional continuity equation:
\begin{equation}
    \frac{\delta\mathcal{A}}
    {\delta\varLambda_{\alpha}}+
    \int_{\mu_{1}}\int_{\mu_{2}}
    \mathcal{K}_{\alpha,\mu_{1}\mu_{2}}
    \left\{\mathcal{A}\,
    \frac{\delta^{2}\mathcal{S}}
    {\delta\varphi_{\mu_{1}}
    \delta\varphi_{\mu_{2}}}+
    2\frac{\delta\mathcal{A}}
    {\delta\varphi_{\mu_{1}}}
    \frac{\delta\mathcal{S}}
    {\delta\varphi_{\mu_{2}}}\right\}=0.
    \label{Continuity_Equation_CN}
\end{equation}

After dividing the imaginary part of the expression (\ref{before_cont_and_HJ}) by $i\mathcal{A}$ we obtain functional HJ equation with a so-called quantum correction term (quantum functional HJ equation):
\begin{equation}
    \frac{\delta\mathcal{S}}
    {\delta\varLambda_{\alpha}}+
    \int_{\mu_{1}}\int_{\mu_{2}}
    \mathcal{K}_{\alpha,\mu_{1}\mu_{2}}
    \frac{\delta\mathcal{S}}
    {\delta\varphi_{\mu_{1}}}
    \frac{\delta\mathcal{S}}
    {\delta\varphi_{\mu_{2}}}=
    \mathcal{U}^{\mathrm{(Quant)}}_{\alpha}+
    \mathcal{U}_{\alpha},
    \label{Quantum_HJ_Equation_CN}
\end{equation}
where the quantum correction term $\mathcal{U}^{\mathrm{(Quant)}}$
\begin{equation}
    \mathcal{U}^{\mathrm{(Quant)}}_{\alpha}=
    \frac{1}{\mathcal{A}}
    \int_{\mu_{1}}\int_{\mu_{2}}
    \mathcal{K}_{\alpha,\mu_{1}\mu_{2}}
    \frac{\delta^{2}\mathcal{A}}
    {\delta\varphi_{\mu_{1}}
    \delta\varphi_{\mu_{2}}}.
    \label{Quantum_Correction_CN}
\end{equation}

Introducing reparametrization of the positive functional $\mathcal{A}$: Let $\mathcal{A}\left[\varLambda, \varphi\right]:=e^{\mathcal{R}\left[\varLambda,\varphi\right]}$. Since the functional $\mathcal{A}$ is the generating functional of the total GFs, and the functional $\mathcal{R}$ is the generating functional of the connected GFs, this reparametrization is natural. Then (dots mean higher-order variational derivatives)
\begin{equation}
  \frac{\delta\mathcal{A}}
  {\delta\varLambda_{\alpha}}=
  e^{\mathcal{R}}\frac{\delta\mathcal{R}}
  {\delta\varLambda_{\alpha}}=
  \mathcal{A}\,\frac{\delta\mathcal{R}}
  {\delta\varLambda_{\alpha}}\ldots
\end{equation}

Dividing the continuity equation (\ref{Continuity_Equation_CN}) by $\mathcal{A}$ yields:
\begin{equation}
    \frac{\delta\mathcal{R}}
    {\delta\varLambda_{\alpha}}+
    \int_{\mu_{1}}\int_{\mu_{2}}
    \mathcal{K}_{\alpha,\mu_{1}\mu_{2}}
    \left\{\frac{\delta^{2}\mathcal{S}}
    {\delta\varphi_{\mu_{1}}
    \delta\varphi_{\mu_{2}}}+
    2\frac{\delta\mathcal{R}}
    {\delta\varphi_{\mu_{1}}}
    \frac{\delta\mathcal{S}}
    {\delta\varphi_{\mu_{2}}}\right\}=0.
    \label{Continuity_Equation_CGFF_CN}
\end{equation}

In order to arrive to the semiclassical approximation for the functional Schr\"{o}dinger equation consider scaling ($\hbar$ is the formal ``reduced Planck constant''): 
\begin{equation}
    \mathcal{S}\rightarrow
    \frac{\mathcal{S}}{\hbar},\quad \mathcal{K}\rightarrow
    \hbar^{2}\mathcal{K},\quad \frac{\delta}{\delta\varLambda}\rightarrow
    \hbar\frac{\delta}{\delta\varLambda}.
    \label{Scaling_Semiclassics}
\end{equation}

The quantum functional HJ equation is not invariant under this transformation. Quantum correction term scales proportional to $\hbar^{2}$:
\begin{equation}
    \mathcal{U}^{\mathrm{(Quant)}}
    \rightarrow\hbar^{2}
    \mathcal{U}^{\mathrm{(Quant)}}.
    \label{Quantum_Correction_Scaling_CN}
\end{equation}

The expression (\ref{Quantum_Correction_Scaling_CN}) implies that in the limit $\hbar\rightarrow 0$ (semiclassical approximation in QFT), the quantum functional HJ equation transforms into the classical one as the quantum correction term vanishes:
\begin{equation}
    \frac{\delta\mathcal{S}}
    {\delta\varLambda_{\alpha}}+
    \int_{\mu_{1}}\int_{\mu_{2}}
    \mathcal{K}_{\alpha,\mu_{1}\mu_{2}}
    \frac{\delta\mathcal{S}}
    {\delta\varphi_{\mu_{1}}}
    \frac{\delta\mathcal{S}}
    {\delta\varphi_{\mu_{2}}}=
    \mathcal{U}_{\alpha}.
    \label{Classical_HJ_Equation_CN}
\end{equation}

\subsection{Hamilton–Jacobi and Continuity Functional Equations Hierarchies}

Let us derive first three equations for GFs corresponding to the functional HJ equation hierarchy. To do so, we perform the following operations (the ``comma'' means evaluated at $\varphi=0$) in both sides of the expression (\ref{Classical_HJ_Equation_CN}):
\begin{equation}
    \mathrm{Eq.1)}\quad\varphi=0;\quad
    \mathrm{Eq.2)}\quad\frac{\delta^{2}}
    {\delta\varphi_{\nu_{1}}\delta\varphi_{\nu_{2}}}, \,\,\varphi=0;\quad
    \mathrm{Eq.3)}\quad\frac{\delta^{4}}
    {\delta\varphi_{\nu_{1}}\delta\varphi_{\nu_{2}}
    \delta\varphi_{\nu_{3}}\delta\varphi_{\nu_{4}}},
    \,\,\varphi=0.
\end{equation}

Odd derivatives would yield zeros since we assume all the functionals to be even with respect to the field $\varphi$. In QM, a solution of the functional Schr\"{o}dinger equation that is even with respect to the spatial coordinate $x$ corresponds to the ground state of the system. The Node Theorem states that eigenstates alternate. Thus, we find both the ground state and the excited ones: the second, the fourth, etc. Let us note that the semiclassical approximation is poorly suited for finding the ground state, but we can use it as the first step of the iterative procedure for solving the exact functional Schr\"{o}dinger equation. Omitting in notation $\left[\varLambda\right]$ sign for explicit functional dependence on $\varLambda$ further in this subsection, we obtain the following equations for GFs:

\begin{equation}
    \frac{\delta^{n}\mathcal{S}
    \left[\varLambda,\varphi\right]}
    {\delta\varphi_{\nu_{1}}\ldots
    \delta\varphi_{\nu_{n}}}
    \bigg\rvert_{\varphi=0}=
    \mathcal{S}^{(n)}_{\nu_{1}\ldots\nu_{n}}
    \left[\varLambda\right],\quad
    \frac{\delta^{n}\mathcal{R}
    \left[\varLambda,\varphi\right]}
    {\delta\varphi_{\nu_{1}}\ldots
    \delta\varphi_{\nu_{n}}}
    \bigg\rvert_{\varphi=0}=
    \mathcal{R}^{(n)}_{\nu_{1}\ldots\nu_{n}}
    \left[\varLambda\right].
    \label{For_The_Reviewer_2_4}
\end{equation}

Equation $1$):
\begin{equation}
    \frac{\delta\mathcal{S}^{(0)}}
    {\delta\varLambda_{\alpha}}=
    \mathcal{U}^{(0)}_{\alpha}.
    \label{HJ_GF_Hierarchy_CN_1}
\end{equation}

Equation $2$):
\begin{equation}
    \frac{\delta\mathcal{S}^{(2)}_{\nu_{1}\nu_{2}}}
    {\delta\varLambda_{\alpha}}+
    2\int_{\mu_{1}}\int_{\mu_{2}}
    \mathcal{K}_{\alpha,\mu_{1}\mu_{2}}
    \mathcal{S}^{(2)}_{\mu_{1}\nu_{1}}
    \mathcal{S}^{(2)}_{\mu_{2}\nu_{2}}=
    \mathcal{U}^{(2)}_{\alpha,\nu_{1}\nu_{2}}.
    \label{HJ_GF_Hierarchy_CN_2}
\end{equation}

Equation $3$):
\begin{equation}
    \frac{\delta
    \mathcal{S}^{(4)}_{\nu_{1}\nu_{2}\nu_{3}\nu_{4}}}
    {\delta\varLambda_{\alpha}}+
    2\int_{\mu_{1}}\int_{\mu_{2}}
    \mathcal{K}_{\alpha,\mu_{1}\mu_{2}}
    \varUpsilon_{\mu_{1}\mu_{2},
    \nu_{1}\nu_{2}\nu_{3}\nu_{4}}=
    \mathcal{U}^{(4)}_{\alpha,
    \nu_{1}\nu_{2}\nu_{3}\nu_{4}},
    \label{HJ_GF_Hierarchy_CN_3}
\end{equation}
where
\begin{equation}
    \varUpsilon_{\mu_{1}\mu_{2},
    \nu_{1}\nu_{2}\nu_{3}\nu_{4}}=
    \mathcal{S}^{(2)}_{\mu_{1}\nu_{1}}
    \mathcal{S}^{(4)}_{\mu_{2}\nu_{2}\nu_{3}\nu_{4}}+
    \mathcal{S}^{(2)}_{\mu_{1}\nu_{2}}
    \mathcal{S}^{(4)}_{\mu_{2}\nu_{1}\nu_{3}\nu_{4}}+
    \mathcal{S}^{(2)}_{\mu_{1}\nu_{3}}
    \mathcal{S}^{(4)}_{\mu_{2}\nu_{1}\nu_{2}\nu_{4}}+
    \mathcal{S}^{(2)}_{\mu_{1}\nu_{4}}
    \mathcal{S}^{(4)}_{\mu_{2}\nu_{1}\nu_{2}\nu_{3}}.
    \label{HJ_GF_Hierarchy_CN_3_Combination}
\end{equation}

The definition of the GFs is given by the expression (\ref{Green_Function_CN}). Let us note that $\forall n>2$ linear and closed equations for $\mathcal{S}^{(n)}$ are obtained and no $n,n+2$ coupling takes place. So, $\forall n\in\mathbb{N}$ GFs $\mathcal{S}^{(n)}$ can be obtained. We also note that $\varphi$-independent functional $\mathcal{K}$ is considered.

Now, let us obtain first two equations for GFs corresponding to the functional continuity equation hierarchy, and to do so, we should apply the following operations to the equation (\ref{Continuity_Equation_CGFF_CN}):
\begin{equation}
    \mathrm{Eq.1')}\quad\varphi=0;\quad
    \mathrm{Eq.2')}\quad\frac{\delta^{2}}
    {\delta\varphi_{\nu_{1}}\delta\varphi_{\nu_{2}}}, \,\,\varphi=0.
\end{equation}

Thus, we obtain the following equations for GFs (recall that the functional $\mathcal{S}$ is the phase of the functional $\varPsi$, while the functional $\mathcal{R}=\ln{|\varPsi|}$):

Equation $1'$):
\begin{equation}
    \frac{\delta\mathcal{R}^{(0)}}
    {\delta\varLambda_{\alpha}}+
    \int_{\mu_{1}}\int_{\mu_{2}}
    \mathcal{K}_{\alpha,\mu_{1}\mu_{2}}
    \mathcal{S}^{(2)}_{\mu_{1}\mu_{2}}=0.
    \label{Continuity_GF_Hierarchy_CN_1}
\end{equation}

Equation $2'$):
\begin{equation}
    \frac{\delta\mathcal{R}^{(2)}_{\nu_{1}\nu_{2}}}
    {\delta\varLambda_{\alpha}}+
    \int_{\mu_{1}}\int_{\mu_{2}}
    \mathcal{K}_{\alpha,\mu_{1}\mu_{2}}
    \left\{\mathcal{S}^{(4)}_{\mu_{1}\mu_{2}
    \nu_{1}\nu_{2}}+
    2\mathcal{R}^{(2)}_{\mu_{1}\nu_{1}}
    \mathcal{S}^{(2)}_{\mu_{2}\nu_{2}}+
    2\mathcal{R}^{(2)}_{\mu_{1}\nu_{2}}
    \mathcal{S}^{(2)}_{\mu_{2}\nu_{1}}\right\}=0.
    \label{Continuity_GF_Hierarchy_CN_2}
\end{equation}

\subsection{Translation-Invariant Solution for Green Functions on Delta-Field Configuration}

In this subsection we present exact translation-invariant solutions first for the equations (\ref{HJ_GF_Hierarchy_CN_1})--(\ref{HJ_GF_Hierarchy_CN_3}) and then for the equations (\ref{Continuity_GF_Hierarchy_CN_1})--(\ref{Continuity_GF_Hierarchy_CN_2}). Let us note that in QFT and Theory of Critical Phenomena, the translation invariance of space (spacetime) is the most natural symmetry since this invariance leads to the momentum (momentum-energy) conservation law in theory, for example, in Feynman diagrams. To obtain exact translation-invariant solutions we choose the delta-field configuration of the holographic field $\varLambda$.

\subsubsection{Hamilton–Jacobi Functional Equation Hierarchy}

For the translation-invariant problem equation (\ref{HJ_GF_Hierarchy_CN_1}) remains unchanged:
\begin{equation}
    \frac{\delta\mathcal{S}^{(0)}
    \left[\varLambda\right]}
    {\delta\varLambda\left(z\right)}=
    \mathcal{U}^{(0)}\left[\varLambda\right]
    \left(z\right).
    \label{HJ_GF_Hierarchy_CN_1_Lambda}
\end{equation}

As before, consider delta-field configuration of $\varLambda$ (\ref{Delta-field_configuration}). Recall that using the reconstruction formula for functionals (\ref{The_reconstruction_formula1}), we made sure that the solutions and conclusions obtained for the delta-field configuration of $\varLambda$ remain valid for a wide class of field configurations. For the delta-field configuration of $\varLambda$ putting $z=w$ the expression (\ref{HJ_GF_Hierarchy_CN_1_Lambda}) transforms as follows:
\begin{equation}
    \frac{\partial\mathcal{S}^{(0)}
    \left(\varLambda;w\right)}
    {\partial\varLambda}=
    \mathcal{U}^{(0)}\left(\varLambda;w\right).
    \label{HJ_GF_Hierarchy_CN_1_Lambda_Delta}
\end{equation}

The general solution of the equation (\ref{HJ_GF_Hierarchy_CN_1_Lambda_Delta}) reads:
\begin{equation}
    \mathcal{S}^{(0)}\left(\varLambda;w\right)=
    \int_{\varLambda_{0}}^{\varLambda}
    d\varLambda'\,\mathcal{U}^{(0)}
    \left(\varLambda';w\right)+
    \mathcal{S}^{(0)}\left(\varLambda_{0};w\right).
    \label{HJ_GF_Hierarchy_CN_1_Lambda_Sol}
\end{equation}

For the equations of the hierarchy being considered of order more then zero (the vacuum mean equation), the condition of translation invariance modifies the equations. For a GF of more than one argument this condition is best visible in the momentum representation. So, performing the Fourier transform of the translation-invariant $n$-particle GF $\mathcal{S}^{(n)}$ in momentum compact notation (\ref{Momentum_Compact_Notation}) we get:
\begin{equation}
\begin{split}
    \mathcal{S}^{(n)}\left[\varLambda\right]
    \left(y_{1},\ldots,y_{n}\right)&=
    \int_{k_{1}}\ldots\int_{k_{n}}
    e^{i\left(k_{1}y_{1}+\ldots+k_{n}y_{n}\right)} 
    \tilde{\mathcal{S}}^{(n)}
    \left[\varLambda\right]
    \left(k_{1},\ldots,k_{n}\right),\\
    \tilde{\mathcal{S}}^{(n)}
    \left[\varLambda\right]
    \left(k_{1},\ldots,k_{n}\right)&=
    (2\pi)^{D}\delta^{(D)}
    \left(k_{1}+\ldots+k_{n}\right)
    \mathcal{S}^{(n)}\left[\varLambda\right]
    \left(k_{1},\ldots,k_{n}\right).
    \label{Sn_Green_Function_TIP_SE}
\end{split}
\end{equation}

The same for the translation-invariant $n$-particle GF $\mathcal{R}^{(n)}$:
\begin{equation}
\begin{split}
    \mathcal{R}^{(n)}\left[\varLambda\right]
    \left(y_{1},\ldots,y_{n}\right)&=
    \int_{k_{1}}\ldots\int_{k_{n}}
    e^{i\left(k_{1}y_{1}+\ldots+k_{n}y_{n}\right)} 
    \tilde{\mathcal{R}}^{(n)}
    \left[\varLambda\right]
    \left(k_{1},\ldots,k_{n}\right),\\
    \tilde{\mathcal{R}}^{(n)}
    \left[\varLambda\right]
    \left(k_{1},\ldots,k_{n}\right)&=
    (2\pi)^{D}\delta^{(D)}
    \left(k_{1}+\ldots+k_{n}\right)
    \mathcal{R}^{(n)}\left[\varLambda\right]
    \left(k_{1},\ldots,k_{n}\right).
    \label{Rn_Green_Function_TIP_SE}
\end{split}
\end{equation}

For the expressions (\ref{HJ_GF_Hierarchy_CN_3})--(\ref{HJ_GF_Hierarchy_CN_3_Combination}) with a condition of translation invariance after simple 
redefinition in momentum representation we obtain:
\begin{equation}
    \frac{\delta\mathcal{S}^{(4)}
    \left[\varLambda\right]
    \left(k_{1},\ldots,k_{4}\right)}
    {\delta\varLambda\left(z\right)}+
    2\varUpsilon\left[\varLambda\right]
    \left(z;k_{1},\ldots,k_{4}\right)=
    \mathcal{U}^{(4)}\left[\varLambda\right]
    \left(z;k_{1},\ldots,k_{4}\right),
    \label{HJ_GF_Hierarchy_CN_3_Lambda}
\end{equation}
where
\begin{equation}
    \varUpsilon\left[\varLambda\right]
    \left(z;k_{1},\ldots,k_{4}\right)=
    \mathcal{S}^{(4)}
    \left[\varLambda\right]
    \left(k_{1},\ldots,k_{4}\right)
    \sum_{j=1}^{4}\mathcal{K}
    \left[\varLambda\right]
    \left(z;k_{j}\right)
    \mathcal{S}^{(2)}
    \left[\varLambda\right]
    \left(-k_{j}\right).
    \label{HJ_GF_Hierarchy_CN_3_Combination_Lambda}
\end{equation}

Considering delta-field configuration of $\varLambda$ (\ref{Delta-field_configuration}) and setting $z=w$ afterwards, the expressions (\ref{HJ_GF_Hierarchy_CN_3_Lambda})--(\ref{HJ_GF_Hierarchy_CN_3_Combination_Lambda}) are converted to the form:
\begin{equation}
    \frac{\partial\mathcal{S}^{(4)}
    \left(\varLambda;w;k_{1},\ldots,k_{4}\right)}
    {\partial\varLambda}+2\varUpsilon
    \left(\varLambda;w;k_{1},\ldots,k_{4}\right)=
    \mathcal{U}^{(4)}
    \left(\varLambda;w;k_{1},\ldots,k_{4}\right),
    \label{HJ_GF_Hierarchy_CN_3_Lambda_Delta}
\end{equation}
where
\begin{equation}
    \varUpsilon\left(\varLambda;w;
    k_{1},\ldots,k_{4}\right)=
    \mathcal{S}^{(4)}\left(\varLambda;w;
    k_{1},\ldots,k_{4}\right)
    \sum_{j=1}^{4}\mathcal{K}
    \left(\varLambda;w;k_{j}\right)
    \mathcal{S}^{(2)}
    \left(\varLambda;w;-k_{j}\right).
    \label{HJ_GF_Hierarchy_CN_3_Combination_Lam_Del}
\end{equation}

In all the expressions (\ref{Sn_Green_Function_TIP_SE})--(\ref{HJ_GF_Hierarchy_CN_3_Combination_Lam_Del}), there is a special kinematics that satisfies the momentum conservation law: $k_{1}+\ldots+k_{4}=0$. The resulting equation (\ref{HJ_GF_Hierarchy_CN_3_Lambda_Delta})--(\ref{HJ_GF_Hierarchy_CN_3_Combination_Lam_Del}) is the linear non-homogeneous differential equation of the first order. Indeed, on the right-hand side of the equation (\ref{HJ_GF_Hierarchy_CN_3_Lambda_Delta}) there is a given function $\mathcal{U}^{(4)}$ of the variable $\varLambda$, while on the left-hand side of the equation (\ref{HJ_GF_Hierarchy_CN_3_Lambda_Delta}) there are the derivative of the unknown function $\mathcal{S}^{(4)}$ with respect to $\varLambda$ and the product $\varUpsilon$ (\ref{HJ_GF_Hierarchy_CN_3_Combination_Lam_Del}) of the unknown function $\mathcal{S}^{(4)}$ and the function $\mathcal{K}\mathcal{S}^{(2)}$ known from the previous hierarchy equation. The solution to such an equation can be obtained in the general form. This solution can be found in any handbook on differential equations. Let us note that linear non-homogeneous differential equations can also be obtained for higher-order GFs $\mathcal{S}^{(n)}$. Only the two-particle GF $\mathcal{S}^{(2)}$, which is the coefficient function of the equation (\ref{HJ_GF_Hierarchy_CN_3_Lambda_Delta})--(\ref{HJ_GF_Hierarchy_CN_3_Combination_Lam_Del}), satisfies the nonlinear equation (\ref{HJ_GF_Hierarchy_CN_2}). As for equation (\ref{HJ_GF_Hierarchy_CN_2}), this equation has already been considered in the Functional Hamilton--Jacobi Equation and its Hierarchy section. Thus, all the GFs $\mathcal{S}^{(n)}$ can be obtained in quadratures. In the next subsubsection, the same situation arises for the GFs $\mathcal{R}^{(n)}$.

\subsubsection{Continuity Functional Equation Hierarchy and Optical Potential}

We are now considering the hierarchy corresponding to the functional continuity equation. Taking into consideration translation invariance condition doesn't change equation (\ref{Continuity_GF_Hierarchy_CN_1}), but creates an infinite term: a vacuum loop. The fact that the vacuum mean is proportional to $(2\pi)^{D}\delta^{(D)}\left(0\right)$ is standard in QFT \cite{nedelko1995oscillator,bogoliubov1980introduction,bogolyubov1983quantum,bogolyubov1990GeneralQFT}. Let us note that the vacuum mean is only in some sense analogous to the vacuum energy density computed with no cutoff at the Planck length, since the wave functional $\varPsi$ in general case doesn't coincide with the scattering matrix. However, in the presented paper, we are interested in obtaining \emph{completely finite wave functional} $\varPsi$. For this reason, we choose the following strategy for solving the functional Schr\"{o}dinger equation: To subtract the diverging vacuum loop, we use the renormalization through so-called (complex-valued) optical potential:
\begin{equation}
    \mathcal{U}\left[\varLambda,\varphi\right]
    \left(z\right)\rightarrow
    \mathcal{U}^{\mathrm{(Optic)}}
    \left[\varLambda,\varphi\right]
    \left(z\right)=\mathcal{U}
    \left[\varLambda,\varphi\right]
    \left(z\right)-i\mathcal{W}
    \left[\varLambda,\varphi\right]
    \left(z\right).
    \label{Optical_Potential}
\end{equation}

Analogously to the optical potential in ordinary QM, the optical potential $\mathcal{U}^{\mathrm{(Optic)}}$ in QFT contains information about the vacuum of the theory, in other words, about (virtual) many-particle vacuum structure. Introducing the optical potential $\mathcal{U}^{\mathrm{(Optic)}}$ into the translation invariant version of the general equation (\ref{Continuity_GF_Hierarchy_CN_1}) yields:
\begin{equation}
    \frac{\delta\mathcal{R}^{(0)}
    \left[\varLambda\right]}
    {\delta\varLambda\left(z\right)}+
    (2\pi)^{D}\delta^{(D)}\left(0\right)
    \int_{k}\mathcal{K}\left[\varLambda\right]
    \left(z;k\right)\mathcal{S}^{(2)}
    \left[\varLambda\right]\left(-k\right)=
    \mathcal{W}\left[\varLambda\right]
    \left(z\right).
    \label{Continuity_GF_Hierarchy_CN_1_Lambda}
\end{equation}

Splitting $\mathcal{W}$ into a sum of singular and regular parts $\mathcal{W}=\mathcal{W}_{\mathrm{sing}}+
\mathcal{W}_{\mathrm{reg}}$ and absorbing an infinite vacuum loop into the singular part $\mathcal{W}_{\mathrm{sing}}$ finishes the regularization:
\begin{equation}
    \frac{\delta\mathcal{R}^{(0)}
    \left[\varLambda\right]}
    {\delta\varLambda\left(z\right)}=
    \mathcal{W}_{\mathrm{reg}}
    \left[\varLambda\right]\left(z\right).
    \label{Continuity_GF_Hierarchy_CN_1_Lambda_Reg}
\end{equation}

Let us note that this choice makes the functional Schr\"{o}dinger equation \emph{nonlinear with respect to the wave functional} $\varPsi$. Indeed, according to the expressions (\ref{Continuity_GF_Hierarchy_CN_1_Lambda})--(\ref{Continuity_GF_Hierarchy_CN_1_Lambda_Reg}), the singular part of the optical potential $\mathcal{W}_{\mathrm{sing}}$ is equal to the integral of the function $\mathcal{S}^{(2)}$. The function $\mathcal{S}^{(2)}$, in turn, is the coefficient of the functional Taylor series expansion of the functional $\mathcal{S}$. The latter is the phase of the functional $\varPsi$.

This situation in a sense repeats a similar one in QFT: The bare parameters of the model are renormalized in accordance with the divergences generated by the integrals of the GFs. Further, considering delta-field configuration of $\varLambda$ (\ref{Delta-field_configuration}) and setting $z=w$ afterwards, the expression (\ref{Continuity_GF_Hierarchy_CN_1_Lambda_Reg}) is converted to the form:
\begin{equation}
    \frac{\partial\mathcal{R}^{(0)}
    \left(\varLambda;w\right)}
    {\partial\varLambda}=
    \mathcal{W}_{\mathrm{reg}}
    \left(\varLambda;w\right).
    \label{Continuity_GF_Hierarchy_CN_1_Lam_Reg_Del}
\end{equation}

The general solution of the equation (\ref{Continuity_GF_Hierarchy_CN_1_Lam_Reg_Del}) reads:
\begin{equation}
    \mathcal{R}^{(0)}\left(\varLambda;w\right)=
    \int_{\varLambda_{0}}^{\varLambda}
    d\varLambda'\,\mathcal{W}_{\mathrm{reg}}
    \left(\varLambda';w\right)+
    \mathcal{R}^{(0)}\left(\varLambda_{0};w\right).
    \label{Continuity_GF_Hierarchy_CN_1_Lam_Reg_Sol}
\end{equation}

Next, for the translation-invariant problem the general equation (\ref{Continuity_GF_Hierarchy_CN_2}) in momentum representation read as follows:
\begin{equation}
    \frac{\delta\mathcal{R}^{(2)} \left[\varLambda\right]\left(k\right)}
    {\delta\varLambda\left(z\right)}+
    \int_{q}\mathcal{K}\left[\varLambda\right]
    \left(z;q\right)\mathcal{S}^{(4)}
    \left[\varLambda\right]
    \left(q,-q,k,-k\right)+
    2\varSigma\left[\varLambda\right]
    \left(z;k\right)=0,
    \label{Continuity_GF_Hierarchy_CN_2_Lambda}
\end{equation}
where the functional $\varSigma$
\begin{equation}
    \varSigma\left[\varLambda\right]
    \left(z;k\right)=\mathcal{K}
    \left[\varLambda\right]\left(z;k\right)
    \mathcal{S}^{(2)} \left[\varLambda\right]\left(k\right)
    \mathcal{R}^{(2)} \left[\varLambda\right]\left(-k\right)+
    \left(k\rightleftarrows -k\right).
    \label{Building_Block_Sigma}
\end{equation}

For the delta-field configuration of $\varLambda$ (\ref{Delta-field_configuration}) and if $z=w$, expressions (\ref{Continuity_GF_Hierarchy_CN_2_Lambda})--(\ref{Building_Block_Sigma}) are converted to the following form:
\begin{equation}
    \frac{\partial\mathcal{R}^{(2)} \left(\varLambda;w;k\right)}
    {\partial\varLambda}+
    \int_{q}\mathcal{K}
    \left(\varLambda;w;q\right)\mathcal{S}^{(4)}
    \left(\varLambda;w;q,-q,k,-k\right)+
    2\varSigma\left(\varLambda;w;k\right)=0,
    \label{Continuity_GF_Hierarchy_CN_2_Lambda_Delta}
\end{equation}
where the function $\varSigma$
\begin{equation}
    \varSigma\left(\varLambda;w;k\right)
    =\mathcal{K}\left(\varLambda;w;k\right)
    \mathcal{S}^{(2)}\left(\varLambda;w;k\right)
    \mathcal{R}^{(2)}\left(\varLambda;w;-k\right)+
    \left(k\rightleftarrows -k\right).
    \label{Building_Block_Sigma_Delta}
\end{equation}

Supposed every function is even:
\begin{equation}
    \varSigma\left(\varLambda;w;k\right)
    =2\mathcal{K}\left(\varLambda;w;k\right)
    \mathcal{S}^{(2)}\left(\varLambda;w;k\right)
    \mathcal{R}^{(2)}\left(\varLambda;w;k\right).
    \label{Building_Block_Sigma_Delta_Even}
\end{equation}

The equation (\ref{Continuity_GF_Hierarchy_CN_2_Lambda_Delta}) for the two-particle GF $\mathcal{R}^{(2)}$ is the linear non-homogeneous differential equation of the first order. The solution to such an equation can be found in any handbook on differential equations. Let us note that linear non-homogeneous differential equations can also be obtained for higher-order GFs $\mathcal{R}^{(n)}$. Thus the problem of hierarchies is completely solvable. In the conclusion of this subsubsection, let us make one more important note: The solution of the functional Schr\"{o}dinger equation can be obtained by the iteration method, where the first step of the iteration is the semiclassical solution.

\subsubsection{Open Quantum Field Systems}

Let us now apply the previously derived integration (reconstruction) formula for functionals (\ref{The_reconstruction_formula1}) to the equation (\ref{Continuity_GF_Hierarchy_CN_1_Lambda}), the second equation of the continuity functional equation hierarchy:
\begin{equation}
    \mathcal{R}^{(0)}\left[\varLambda+
    \varDelta\right]-\mathcal{R}^{(0)}
    \left[\varLambda\right]+
    \left(2\pi\right)^{D}\delta^{(D)}
    \left(0\right)\mathrm{Loop}
    \left[\varLambda;\varDelta\right]=
    \int\limits_{0}^{1}dt
    \int d^{D}z\,\varDelta\left(z\right)
    \mathcal{W}\left[\varLambda+
    t\varDelta\right]\left(z\right),
    \label{Open_QFT1}
\end{equation}
where the functional $\mathrm{Loop}$
\begin{equation}
    \mathrm{Loop}\left[\varLambda;
    \varDelta\right]\equiv
    \int\limits_{0}^{1}dt
    \int d^{D}z\,\varDelta\left(z\right)
    \int_{k}\mathcal{K}\left[\varLambda+
    t\varDelta\right]\left(z;k\right)
    \mathcal{S}^{(2)}\left[\varLambda+
    t\varDelta\right]\left(-k\right).
    \label{Open_QFT2}
\end{equation}

We see that the singular term proportional to $(2\pi)^{D}\delta^{(D)}\left(0\right)$ arises for any configurations of the field $\varLambda$. For completeness of the previous results, consider the opposite to the delta-field case (the constant-field configurations): $\varLambda\left(z\right)=\varLambda$ and $\varDelta\left(z\right)=\varDelta$. As before, set $\varLambda=\varLambda_{0}$, $\varDelta=\varLambda_{1}-\varLambda_{0}$ and $t\varDelta=\varLambda_{t}-\varLambda_{0}$. The expressions (\ref{Open_QFT1})--(\ref{Open_QFT2}) are simplified as follows:
\begin{equation}
    \mathcal{R}^{(0)}\left(\varLambda_{1}
    \right)-\mathcal{R}^{(0)}\left(\varLambda_{0}
    \right)+\left(2\pi\right)^{D}\delta^{(D)}
    \left(0\right)\mathrm{Loop}
    \left(\varLambda_{1};\varLambda_{0}\right)=
    \int\limits_{\varLambda_{0}}^{\varLambda_{1}}
    d\varLambda_{t}
    \int d^{D}z\,
    \mathcal{W}\left(\varLambda_{t};z\right),
    \label{Open_QFT3}
\end{equation}
where the function $\mathrm{Loop}$
\begin{equation}
    \mathrm{Loop}
    \left(\varLambda_{1};\varLambda_{0}\right)\equiv
    \int\limits_{\varLambda_{0}}^{\varLambda_{1}}
    d\varLambda_{t}\int d^{D}z\,
    \int_{k}\mathcal{K}\left(\varLambda_{t};z;k\right)
    \mathcal{S}^{(2)}\left(\varLambda_{t};-k\right).
    \label{Open_QFT4}
\end{equation}

As before, we are interested in obtaining completely finite wave functional $\varPsi$. For this reason, we choose the following regularization:
\begin{equation}
    \int\limits_{\varLambda_{0}}^{\varLambda_{1}}
    d\varLambda_{t}
    \int d^{D}z\,
    \mathcal{W}_{\mathrm{sing}}
    \left(\varLambda_{t};z\right)=
    \left(2\pi\right)^{D}\delta^{(D)}
    \left(0\right)\mathrm{Loop}
    \left(\varLambda_{1};\varLambda_{0}\right).
    \label{Open_QFT5}
\end{equation}

This choice again makes the functional Schr\"{o}dinger equation nonlinear with respect to the wave functional $\varPsi$. Thus, the singular part of the optical potential $\mathcal{W}_{\mathrm{sing}}$ corresponds to the dynamics of the quantum field system vacuum, while the regular part of the optical potential $\mathcal{W}_{\mathrm{reg}}$ corresponds to a controlled external action on the quantum field system. At the end of the section, we make the following remarks. First, if we abandon the requirement that the vacuum mean $\mathcal{R}^{(0)}$ is finite and define the function $r$
\begin{equation}
    \mathcal{R}^{(0)}\left(\varLambda\right)=
    \left(2\pi\right)^{D}\delta^{(D)}
    \left(0\right)r\left(\varLambda\right),
    \label{Open_QFT6}
\end{equation}
then we can assume $\mathcal{W}$ =0 (no optical potential is required). The equation for the function $r$ reads:
\begin{equation}
    r\left(\varLambda_{1}
    \right)-r\left(\varLambda_{0}
    \right)+\mathrm{Loop}
    \left(\varLambda_{1};\varLambda_{0}\right)=0.
    \label{Open_QFT7}
\end{equation}

Second, if the functional Hamiltonian $\hat{\mathcal{H}}$ is a non-Hermitian operator, then the eigenvalues can be from $\mathbb{R}$. In general, however, the translation-invariant functional Schr\"{o}dinger equation describes an open quantum field system. The presence of the translation invariance is important: If we abandon the latter, we can obtain a finite expression for the vacuum mean $\mathcal{R}^{(0)}$.

\section{Wilson--Polchinski Functional Equation and Functional Renormalization Group}

In this section, we apply experience gained in two previous sections to the study of the FRG flow equations \cite{kopbarsch,wipf2012statistical,rosten2012fundamentals,igarashi2009realization}. In particular, we establish an analogue of the semiclassical approximation for the WP functional equation. In this approximation, the latter is reduced to the functional HJ equation. We give a detailed solution to the resulting equation. Also we investigate in general terms the question of when the modes coarse graining growth functional can be considered purely geometric, in other words, when does it depend only on the functional momentum, but not on the derivatives of the momentum.

\subsection{Quantum and Classical Parts of the Wilson–Polchinski Equation}

Let us consider the ambiguity of the FRG flow procedures (with some insignificant differences in the definition of the modes coarse graining growth functional in comparison with the review \cite{rosten2012fundamentals}). For definiteness the WP functional equation for the generating functional $\mathcal{G}$ of the amputated GFs is considered. This equation is a special case of a more general functional flow construction which can be formulated in a form of implicit (with respect to the functional $\mathcal{G}$) functional (master) equation. In the coordinate representation this master equation is:
\begin{equation}
    \frac{\delta\mathcal{G}
    \left[\varLambda,\varphi\right]}
    {\delta\varLambda\left(z\right)}=
    \int d^{D}x\,
    \frac{\delta\varPsi
    \left[\varLambda,\varphi,\varpi\right]
    \left(z;x\right)}
    {\delta\varphi\left(x\right)}+
    \int d^{D}x\,
    \varPsi\left[\varLambda,\varphi,\varpi\right]
    \left(z;x\right)
    \varpi\left(x\right),\quad
    \varpi\left(x\right)=\frac{\delta\mathcal{G}
    \left[\varLambda,\varphi\right]}
    {\delta\varphi\left(x\right)}.
    \label{ch4-eq5}
\end{equation}

In the condensed notation this master equation is:
\begin{equation}
    \frac{\delta\mathcal{G}
    \left[\varLambda,\varphi\right]}
    {\delta\varLambda_{\alpha}}=
    \int_{\mu}
    \frac{\delta\varPsi_{\alpha,\mu}
    \left[\varLambda,\varphi,\varpi\right]}
    {\delta\varphi_{\mu}}+
    \int_{\mu}\varPsi_{\alpha,\mu}
    \left[\varLambda,\varphi,\varpi\right]
    \varpi_{\mu},\quad
    \varpi_{\mu}=\frac{\delta\mathcal{G}
    \left[\varLambda,\varphi\right]}
    {\delta\varphi_{\mu}}.
    \label{ch4-eq6}
\end{equation}

The object $\varPsi$ has one ``index'' $x$ and is a functional of a field variable $\varphi$. This functional is the modes coarse graining growth functional in FRG. Moreover, it depends on the functional $\mathcal{G}$ which makes the functional flow equations (\ref{ch4-eq5})--(\ref{ch4-eq6}) implicit. The meaning of $\varPsi$ is that this object parameterizes the process of increasing of coarse grain level for degrees of freedom in the system, in other words, one or another FRG flow procedure. At the same moment $\varPsi$ satisfies only general conditions and its specific form is up to a particular case. Let us also make an important note: We assume that $\varPsi$ depends only on the functional momentum $\varpi$, and doesn't depend on derivatives of $\varpi$ with respect to $\varphi$. Such a functional can be called ``geometric''. The validity of this assumption will be shown later in the section.

In order to get the WP functional equation from (\ref{ch4-eq5})--(\ref{ch4-eq6}) we need to make the following choice in the coordinate representation:
\begin{equation}
    \varPsi\left[\varLambda,\varphi,\varpi\right]
    \left(z;x\right)=\frac{1}{2}\int d^{D}y\,
    \frac{\delta G\left[\varLambda\right]
    \left(x,y\right)}
    {\delta\varLambda\left(z\right)}\,
    \varpi\left(y\right).
    \label{ch4-eq7}
\end{equation}

In the condensed notation this expression is:
\begin{equation}
    \varPsi_{\alpha,\mu}
    \left[\varLambda,\varphi,\varpi\right]=
    \frac{1}{2}\int_{\nu}
    \frac{\delta G_{\mu\nu}\left[\varLambda\right]}
    {\delta\varLambda_{\alpha}}\,
    \varpi_{\nu}.
    \label{ch4-eq8}
\end{equation}

The integral kernel in the expressions (\ref{ch4-eq7})--(\ref{ch4-eq8}), which is the variational derivative of some functional $G$ ($\varLambda$-deformed propagator) with respect to field $\varLambda$ will be defined in the next subsection. Therefore, the functional equations (\ref{ch4-eq5})--(\ref{ch4-eq6}) demonstrate us a large functional ambiguity of the FRG method. Nevertheless, this ambiguity is not a drawback but an opportunity that may provide us a deeper insight into QFT. Thus, the WP functional equation for the generating functional $\mathcal{G}$ of the amputated GFs is given by \cite{kopbarsch,wipf2012statistical,rosten2012fundamentals,igarashi2009realization}:
\begin{equation}
    \frac{\delta\mathcal{G}
    \left[\varLambda,\varphi\right]}
    {\delta\varLambda\left(z\right)}=
    \frac{1}{2}\int d^{D}x_{1}
    \int d^{D}x_{2}\,
    \frac{\delta G\left[\varLambda\right]
    \left(x_{1},x_{2}\right)}
    {\delta\varLambda\left(z\right)}
    \left\{\frac{\delta^{2}
    \mathcal{G}\left[\varLambda,\varphi\right]}
    {\delta\varphi\left(x_{1}\right)
    \delta\varphi\left(x_{2}\right)}+
    \frac{\delta\mathcal{G}
    \left[\varLambda,\varphi\right]}
    {\delta\varphi\left(x_{1}\right)}
    \frac{\delta\mathcal{G}
    \left[\varLambda,\varphi\right]}
    {\delta\varphi\left(x_{2}\right)}\right\}.
    \label{ch4-eq3}
\end{equation}

In the condensed notation this expression is:
\begin{equation}
    \frac{\delta\mathcal{G}
    \left[\varLambda,\varphi\right]}
    {\delta\varLambda_{\alpha}}=
    \frac{1}{2}\int_{\mu_{1}}
    \int_{\mu_{2}}\,
    \frac{\delta G_{\mu_{1}\mu_{2}}
    \left[\varLambda\right]}
    {\delta\varLambda_{\alpha}}
    \left\{\frac{\delta^{2}
    \mathcal{G}\left[\varLambda,\varphi\right]}
    {\delta\varphi_{\mu_{1}}
    \delta\varphi_{\mu_{2}}}+
    \frac{\delta\mathcal{G}
    \left[\varLambda,\varphi\right]}
    {\delta\varphi_{\mu_{1}}}
    \frac{\delta\mathcal{G}
    \left[\varLambda,\varphi\right]}
    {\delta\varphi_{\mu_{2}}}\right\}.
    \label{ch4-eq4}
\end{equation}

In some literature \cite{kopbarsch}, the functional equations (\ref{ch4-eq3})--(\ref{ch4-eq4}) are called the WP FRG flow equations in the coordinate representation and in the abstract form, correspondingly. For the solution of these equations we can use the expansion of the functional $\mathcal{G}$ in the functional Taylor series with respect to the field configurations $\varphi$. This expansion generates an infinite hierarchy (chain) of coupled integro-differential equations for the corresponding functions $\mathcal{G}^{(n)}$, which are the expansion coefficients for the functional Taylor series. For example, the equation for the two-particle function $\mathcal{G}^{(2)}$ contains also the four-particle function $\mathcal{G}^{(4)}$ in special kinematics (for simplicity we assume that the functions of odd order $\mathcal{G}^{(n)}$ identically equal to zero which is correct for theories with interaction Lagrangians even in the field $\varphi$). Next, the equation for the four-particle function $\mathcal{G}^{(4)}$ contains the six-particle function $\mathcal{G}^{(6)}$ (in special kinematics) and so on. As noted in the Introduction section, this is called ``$n,\,n+2$ problem'', and this problem is the main difficulty in the FRG method.

The generating functional $\mathcal{G}$ of the amputated connected Green functions is given by the negative of the interaction part of known initial action $S_{\mathrm{int}}$ if $\varLambda\rightarrow +\infty$ \cite{kopbarsch}:
\begin{equation}
    \mathcal{G}\left[\varLambda,\varphi\right]
    \rightarrow -S_{\mathrm{int}}\left[\varphi\right]=
    -\int d^{D}x\, g\left(x\right)
    V\left[\varphi\left(x\right)\right],
    \quad\varLambda\rightarrow +\infty.
    \label{ch4-eq11}
\end{equation}

The expression (\ref{ch4-eq11}) is a more convenient starting point for approximations than the zero initial condition. Further, replace the quantum part of the WP functional equation as follows:
\begin{equation}
    \frac{\delta\mathcal{G}
    \left[\varLambda,\varphi\right]}
    {\delta\varLambda\left(z\right)}=
    \frac{1}{2}\int d^{D}x_{1}
    \int d^{D}x_{2}\,
    \frac{\delta G\left[\varLambda\right]
    \left(x_{1},x_{2}\right)}
    {\delta\varLambda\left(z\right)}
    \left\{\frac{\delta\mathcal{G}
    \left[\varLambda,\varphi\right]}
    {\delta\varphi\left(x_{1}\right)}
    \frac{\delta\mathcal{G}
    \left[\varLambda,\varphi\right]}
    {\delta\varphi\left(x_{2}\right)}-
    \frac{\delta^{2}
    S_{\mathrm{int}}\left[\varphi\right]}
    {\delta\varphi\left(x_{1}\right)
    \delta\varphi\left(x_{2}\right)}\right\}.
    \label{ch4-eq12}
\end{equation}

Thus, we arrive at the functional HJ equation of the special form. As noted in the Introduction section, equations of this type are basic in the HRG method \cite{lizana2016holographic,akhmedov1998remark,de2000holographic,verlinde2000rg,fukuma2003holographic,akhmedov2003notes,akhmedov2011hints,heemskerk2011holographic}. In the next two subsections, we obtain the translation-invariant solution to this equation hierarchy. This is possible since the functional equation (\ref{ch4-eq12}) doesn't contain ``$n,\,n+2$ problem''.

\subsection{Solution of the Approximated Wilson--Polchinski Equation: Two-Particle Green Function}

Repeating the calculations presented in Functional Hamilton–Jacobi Equation and its Hierarchy section, it is easy to derive an equation for the two-particle amputated GF $\mathcal{G}^{(2)}$. In general, this equation has the form:
\begin{equation}
\begin{split}
    &\frac{\delta\mathcal{G}^{(2)}
    \left[\varLambda\right]\left(y_{1},y_{2}\right)}
    {\delta\varLambda\left(z\right)}=
    -\frac{1}{2}\int d^{D}x_{1}
    \int d^{D}x_{2}\,
    \frac{\delta G\left[\varLambda\right]
    \left(x_{1},x_{2}\right)}
    {\delta\varLambda\left(z\right)}\,
    S^{(4)}_{\mathrm{int}}
    \left(x_{1},x_{2},y_{1},y_{2}\right)+\\
    &+\int d^{D}x_{1}
    \int d^{D}x_{2}\,
    \frac{\delta G\left[\varLambda\right]
    \left(x_{1},x_{2}\right)}
    {\delta\varLambda\left(z\right)}\,
    \mathcal{G}^{(2)}
    \left[\varLambda\right]\left(x_{1},y_{1}\right)
    \mathcal{G}^{(2)}
    \left[\varLambda\right]\left(x_{2},y_{2}\right).
    \label{ch4-eq13}
\end{split}
\end{equation}

For the translation-invariant problem the function $g(x)=g$ in the expression (\ref{ch4-eq11}). Thus, the equation (\ref{ch4-eq13}) for two-particle GF in momentum representation reads as follows:
\begin{equation}
    \frac{\delta\mathcal{G}^{(2)}
    \left[\varLambda\right]\left(k\right)}
    {\delta\varLambda\left(z\right)}=
    -\frac{1}{2}\int_{k'}\frac{\delta G\left[\varLambda\right]\left(k'\right)}
    {\delta\varLambda\left(z\right)}\,
    S^{(4)}_{\mathrm{int}}
    \left(k',-k',k,-k\right)+
    \frac{\delta G\left[\varLambda\right]
    \left(k\right)}
    {\delta\varLambda\left(z\right)}
    |\mathcal{G}^{(2)}\left[\varLambda\right]
    \left(k\right)|^{2}.
    \label{ch4-eq14}
\end{equation}

Further, for the delta-field configuration of $\varLambda$ (\ref{Delta-field_configuration}) and if $z=w$, the expression (\ref{ch4-eq14}) is converted to the following form:
\begin{equation}
    \frac{\partial\mathcal{G}^{(2)}
    \left(\varLambda;w;k\right)}
    {\partial\varLambda}=
    -\frac{1}{2}\int_{k'}
    \frac{\partial G
    \left(\varLambda;w;k'\right)}
    {\partial\varLambda}\,
    S^{(4)}_{\mathrm{int}}
    \left(k',-k',k,-k\right)+
    \frac{\partial G
    \left(\varLambda;w;k\right)}
    {\partial\varLambda}
    |\mathcal{G}^{(2)}
    \left(\varLambda;w;k\right)|^{2}.
    \label{ch4-eq15}
\end{equation}

In what follows, we omit the $w$-dependence. For the polynomial $\varphi^{4}$ theory the interaction part of initial action $S_{\mathrm{int}}$ (\ref{ch4-eq11}) reads:
\begin{equation}
    S_{\mathrm{int}}\left[\varphi\right]=
    g\int d^{D}x\varphi^{4}\left(x\right),
    \quad S^{(4)}_{\mathrm{int}}
    \left(k',-k',k,-k\right)=g.
    \label{ch4-eq16}
\end{equation}

To go further, it is necessary to specify the integral kernel appearing in the expressions above, starting from (\ref{ch4-eq7})--(\ref{ch4-eq8}). For this reason, let us consider $\varLambda$-deformed propagator $G\left(\varLambda;k\right)$ of the theory under consideration known: The deformation is introduced into the original propagator $G\left(k\right)$ so that the following condition is satisfied:
\begin{equation}
    G\left(k\right)\rightarrow
    G\left(\varLambda;k\right),\quad
    G\left(\varLambda;k\right)\sim
    \begin{cases}
    G\left(k\right)\!\!\! & \textrm{for}\,\,\, \varLambda\rightarrow 0,\\
    0\!\!\! & \textrm{for}\,\,\, 
    \varLambda\rightarrow \infty.
    \end{cases}
    \label{3frgstt2}
\end{equation}

Let us note here that we use the notation of functions, although everything remains valid at the functional level (in the latter case, of course, we have a larger space of possibilities). As it is known \cite{kopbarsch,wipf2012statistical,rosten2012fundamentals,igarashi2009realization}, there are various ways to deform the theory propagator. As an example, one can use the following multiplicative deformation of the propagator with some cutoff function (multiplicative regulator) $\varTheta\left(\varLambda;k\right)$:
\begin{equation}
    G\left(\varLambda;k\right)=
    G\left(k\right)
    \varTheta\left(\varLambda;k\right),\quad
    \varTheta\left(\varLambda;k\right)\sim
    \begin{cases}
    1\!\!\! & \textrm{for}\,\,\, 
    \varLambda\rightarrow 0,\\
    0\!\!\! & \textrm{for}\,\,\, 
    \varLambda\rightarrow \infty.
    \end{cases}
    \label{3frgstt3}
\end{equation}

Another example is the additive deformation of the inverse propagator with some cutoff function (additive regulator) $R\left(\varLambda;k\right)$:
\begin{equation}
    \frac{1}{G\left(\varLambda;k\right)}=
    \frac{1}{G\left(k\right)}+
    R\left(\varLambda;k\right),\quad
    R\left(\varLambda;k\right)\sim
    \begin{cases}
    0\!\!\! & \textrm{for}\,\,\, 
    \varLambda\rightarrow 0,\\
    \infty\!\!\! & \textrm{for}\,\,\, \varLambda\rightarrow\infty.
    \end{cases}
    \label{3frgstt4}
\end{equation}

The multiplicative regulator $\varTheta\left(\varLambda;k\right)$ and additive regulator $R\left(\varLambda;k\right)$ can be expressed through each other using the following expressions:
\begin{equation}
    \varTheta\left(\varLambda;k\right)=
    \frac{1}{1+G\left(k\right)
    R\left(\varLambda;k\right)},\quad
    R\left(\varLambda;k\right)=
    \frac{1}{G\left(k\right)}
    \left(\frac{1}
    {\varTheta\left(\varLambda;k\right)}-1\right).
    \label{3frgstt5}
\end{equation}

In practice, however, the convenience of a particular regulator depends on the problem. It also depends on the problem which function to choose as the regulator itself. In this paper, we use the optimized (Litim) additive regulator \cite{kopbarsch,wipf2012statistical}, because it provides analytical expressions for the $\varLambda$-dependent amputated GF. The optimized (Litim) additive regulator reads as follows:
\begin{equation}
    R\left(\varLambda;k\right)=
    \left(\frac{1}{G\left(\varLambda\right)}-
    \frac{1}{G\left(k\right)}\right)
    \varTheta\left(\varLambda-k\right).
    \label{3frgstt6}
\end{equation}

Let us note that within the framework of the problem considered in the paper, the non-analytical behavior of the regulator, Heaviside step function $\varTheta\left(\varLambda-k\right)$, doesn't lead to any non-physical behavior. Further, for the optimized additive regulator (\ref{3frgstt6}) $\varLambda$-deformed propagator $G\left(\varLambda;k\right)$ is:
\begin{equation}
    G\left(\varLambda;k\right)=
    \frac{G\left(k\right)}{1+
    \left(\frac{G\left(k\right)}
    {G\left(\varLambda\right)}-1\right)
    \varTheta\left(\varLambda-k\right)},\quad
    G\left(\varLambda;k\right)=
    \begin{cases}
    G\left(k\right)\!\!\! & 
    \textrm{for}\,\,\, 
    \varLambda<k,\\
    G\left(\varLambda\right)\!\!\! & 
    \textrm{for}\,\,\, 
    \varLambda>k.
    \end{cases}
    \label{3frgstt7}
\end{equation}

One of the optimized additive regulator (\ref{3frgstt6}) advantages is that the derivative of the $\varLambda$-deformed propagator $G\left(\varLambda;k\right)$ with respect to $\varLambda$ has the simplest form:
\begin{equation}
    \frac{\partial G\left(\varLambda;k\right)}
    {\partial\varLambda}=
    \frac{\partial G\left(\varLambda\right)}
    {\partial\varLambda}\,
    \varTheta\left(\varLambda-k\right).
    \label{3frgstt8}
\end{equation}

The derivative (\ref{3frgstt8}) is the integral kernel that we wanted to get explicitly. Thus, the optimized additive regulator (\ref{3frgstt6}) provides the simplest integral kernel. Thus, starting from expression (\ref{ch4-eq15}), we arrive at the following FRG flow equation for the two-particle amputated GF $\mathcal{G}^{(2)}$, which is the simplest among all the possible regulators:
\begin{equation}
    \frac{\partial\mathcal{G}^{(2)}
    \left(\varLambda;k\right)}
    {\partial\varLambda}=
    \frac{\partial G\left(\varLambda\right)}
    {\partial\varLambda}
    \left\{\varTheta\left(\varLambda-k\right)
    |\mathcal{G}^{(2)}
    \left(\varLambda;k\right)|^{2}
    -\frac{B_{D}(1)g\varLambda^{D}}
    {2\left(2\pi\right)^{2}}\right\}.
    \label{3frgstt9}
\end{equation}

In the right-hand side of the expression (\ref{3frgstt9}) $B_{D}(1)$ is the volume of the $D$-dimensional unit radius ball. Let's make additional definitions and assumptions about the power dependence of the propagator and the parity of the functions ($n\in\mathbb{N}$):
\begin{equation}
    G\left(\varLambda\right)=
    \frac{1}{\varLambda^{2n}},\quad
    g_{D}=\frac{B_{D}(1)g}
    {2\left(2\pi\right)^{2}},\quad
    \mathcal{G}^{(2)}
    \left(\varLambda;-k\right)=
    \mathcal{G}^{(2)}
    \left(\varLambda;k\right)=
    \mathcal{G}\left(\varLambda;k\right).
    \label{3frgstt10}
\end{equation}

Within the expression (\ref{3frgstt10}), the equation (\ref{3frgstt9}) reads as follows:
\begin{equation}
    \frac{\partial\mathcal{G}
    \left(\varLambda;k\right)}
    {\partial\varLambda}=
    \frac{2n}{\varLambda^{2n+1}}
    \left\{-\varTheta\left(\varLambda-k\right)
    \mathcal{G}^{2}
    \left(\varLambda;k\right)
    +g_{D}\varLambda^{D}\right\}.
    \label{3frgstt11}
\end{equation}

Before going any further, there is one important point to make. Renormalizable theories are usually explored in the literature. In this case, the boundary condition for the solution $\mathcal{G}\left(\varLambda;k\right)$ of the equation (\ref{3frgstt11}) is set at infinity in $\varLambda$ according to the expression (\ref{ch4-eq11}). However, in the general case, the theory is nonrenormalizable. The latter leads to the fact that the boundary condition for the solution $\mathcal{G}\left(\varLambda;k\right)$ of the equation (\ref{3frgstt11}) is set at some $\varLambda=\varLambda_{0}$, which is the largest scale in the theory \cite{igarashi2009realization}. If the theory turns out to be renormalizable, there is a meaningful limit $\varLambda_{0}\rightarrow\infty$ for all obtained expressions. For this reason, we give the formulation of the problem with a finite $\varLambda_{0}$, after which we consider the limit $\varLambda_{0}\rightarrow\infty$. When $\varLambda<k<\varLambda_{0}$ the solution $\mathcal{G}\left(\varLambda;k\right)$ of the equation (\ref{3frgstt11}) has the following form:
\begin{equation}
    \mathcal{G}\left(\varLambda;k\right)=
    \frac{2ng_{D}}{D-2n}
    \left(\varLambda^{D-2n}-
    k^{D-2n}\right)+\alpha\left(k\right).
    \label{3frgstt12}
\end{equation}

The ``constant'' of integration $\alpha\left(k\right)$ appearing in the expression (\ref{3frgstt12}) is the function that doesn't depend on $\varLambda$. This function is determined from the equality condition for solutions at the point $\varLambda=k$. When $k<\varLambda<\varLambda_{0}$ the equation (\ref{3frgstt11}) can be transformed as follows:
\begin{equation}
    \frac{\partial\mathcal{G}
    \left(\varLambda;k\right)}
    {\partial\varLambda}=
    \frac{2n}{\varLambda^{2n+1}}
    \left\{-\mathcal{G}^{2}
    \left(\varLambda;k\right)
    +g_{D}\varLambda^{D}\right\}.
    \label{3frgstt13}
\end{equation}

The equation (\ref{3frgstt13}) is a special case of the expressions (\ref{Bessel_3})--(\ref{Bessel_5}). Consider the case $n=1$ and $D=3$. In this case, the polynomial $\varphi^{4}$ theory (\ref{ch4-eq16}) turns out to be renormalizable \cite{zinn1989field,vasil2004field}. However, we derive this from a general solution containing $\varLambda_{0}$. The latter reads as follows:
\begin{equation}
    \mathcal{G}\left(\varLambda;k\right)=
    \mathcal{G}\left(\varLambda\right)=
    \sqrt{g_{3}\varLambda^{3}}\,
    \frac{I_{1}\left(\lambda_{0}\right)
    K_{1}\left(\lambda\right)
    -K_{1}\left(\lambda_{0}\right)
    I_{1}\left(\lambda\right)}
    {I_{1}\left(\lambda_{0}\right)
    K_{2}\left(\lambda\right)+
    K_{1}\left(\lambda_{0}\right)
    I_{2}\left(\lambda\right)},\quad
    \lambda=\frac{4\sqrt{g_{3}}}
    {\sqrt{\varLambda}},\quad
    \lambda_{0}=\frac{4\sqrt{g_{3}}}
    {\sqrt{\varLambda_{0}}}.
    \label{3frgstt14}
\end{equation}

Let us note here that the solution (\ref{3frgstt14}) depends only on $\varLambda$. The equality condition for solutions at the point $\varLambda=k$ gives the following expression for the function $\alpha\left(k\right)$ appearing in the solution (\ref{3frgstt12}):
\begin{equation}
    \alpha\left(k\right)=
    \mathcal{G}\left(k\right)=
    \sqrt{g_{3}k^{3}}\,
    \frac{I_{1}\left(\lambda_{0}\right)
    K_{1}\left(\kappa\right)
    -K_{1}\left(\lambda_{0}\right)
    I_{1}\left(\kappa\right)}
    {I_{1}\left(\lambda_{0}\right)
    K_{2}\left(\kappa\right)+
    K_{1}\left(\lambda_{0}\right)
    I_{2}\left(\kappa\right)},\quad
    \kappa=\frac{4\sqrt{g_{3}}}
    {\sqrt{k}}.
    \label{3frgstt15}
\end{equation}

Now we can rewrite the solution (\ref{3frgstt12}) for $\varLambda<k<\varLambda_{0}$ as:
\begin{equation}
    \mathcal{G}\left(\varLambda;k\right)=
    2g_{3}\left(\varLambda-k\right)+
    \sqrt{g_{3}k^{3}}\,
    \frac{I_{1}\left(\lambda_{0}\right)
    K_{1}\left(\kappa\right)
    -K_{1}\left(\lambda_{0}\right)
    I_{1}\left(\kappa\right)}
    {I_{1}\left(\lambda_{0}\right)
    K_{2}\left(\kappa\right)+
    K_{1}\left(\lambda_{0}\right)
    I_{2}\left(\kappa\right)}.
    \label{3frgstt16}
\end{equation}

Let us also note that the dependence of the solution (\ref{3frgstt16}) on $\varLambda_{0}$ appears through the equality condition for solutions at the point $\varLambda=k$. Further, according to the general principles of the FRG method, the physical $n$-particle amputated GFs are the values of the $\varLambda$-dependent $n$-particle amputated GFs at $\varLambda=0$. This statement is a consequence of the fact that the $\varLambda$-dependent generating functional $\mathcal{G}$ of amputated GFs of is equal to its physical (quantum) analog at $\varLambda=0$. Thus, for the physical two-particle amputated GF $\mathcal{G}^{\mathrm{(Quant)}}$ we have the following expression:
\begin{equation}
    \mathcal{G}\left(\varLambda=0;k\right)=
    \mathcal{G}^{\mathrm{(Quant)}}\left(k\right)=
    -2g_{3}k+\sqrt{g_{3}k^{3}}\,
    \frac{I_{1}\left(\lambda_{0}\right)
    K_{1}\left(\kappa\right)
    -K_{1}\left(\lambda_{0}\right)
    I_{1}\left(\kappa\right)}
    {I_{1}\left(\lambda_{0}\right)
    K_{2}\left(\kappa\right)+
    K_{1}\left(\lambda_{0}\right)
    I_{2}\left(\kappa\right)}.
    \label{3frgstt17}
\end{equation}

Calculating the limit $\varLambda_{0}\rightarrow\infty$ of the expression (\ref{3frgstt17}), we arrive at the following result:
\begin{equation}
    \mathcal{G}\left(\varLambda=0;k\right)=
    \mathcal{G}^{\mathrm{(Quant)}}\left(k\right)=
    -2g_{3}k+\sqrt{g_{3}k^{3}}\,
    \frac{K_{1}\left(\kappa\right)}
    {K_{2}\left(\kappa\right)}.
    \label{3frgstt18}
\end{equation}

The existence of such a limit is an independent verification of our calculations. The theory under consideration, as it should be, turns out to be renormalizable. Traveling back, let us make an important remark: The solution of the equation (\ref{3frgstt13}) from the domain $k<\varLambda<\varLambda_{0}$ ``knows'' about the pole $D-4$, and about $\ln\varLambda$-dependence at $D=4$. This is a well-established fact from Theory of Critical Phenomena \cite{zinn1989field,vasil2004field}. Such a solution can be written explicitly in terms of Bessel functions. Thus, the semiclassical approximation for the WP functional equation gives reliable results. These results, of course, can be improved by taking several subsequent steps of the iterative solution of the WP functional equation, similar to the iterative solution of the functional Schr\"{o}dinger equation, where the first step of the iteration is also the semiclassics.

\subsection{Solution of the Approximated Wilson--Polchinski Equation: Four-Particle Green Function}

Repeating the calculations presented in Functional Hamilton–Jacobi Equation and its Hierarchy section, it is easy to derive an equation for the four-particle amputated GF $\mathcal{G}^{(4)}$. In general, this equation has the form:
\begin{equation}
    \frac{\delta\mathcal{G}^{(4)}
    \left[\varLambda\right]
    \left(y_{1},y_{2},y_{3},y_{4}\right)}
    {\delta\varLambda\left(z\right)}=
    \int d^{D}x_{1}
    \int d^{D}x_{2}\,
    \frac{\delta G\left[\varLambda\right]
    \left(x_{1},x_{2}\right)}
    {\delta\varLambda\left(z\right)}\,
    \varUpsilon\left[\varLambda\right]
    \left(x_{1},x_{2};y_{1},y_{2},y_{3},y_{4}\right).
    \label{four_particle_function_FRG1}
\end{equation}

Omitting in notation $\left[\varLambda\right]$ sign for explicit functional dependence on $\varLambda$, the functional $\varUpsilon$ is:
\begin{equation}
    \varUpsilon
    \left(x_{1},x_{2};y_{1},y_{2},y_{3},y_{4}\right)=
    \mathcal{G}^{(2)}\left(x_{1},y_{1}\right)
    \mathcal{G}^{(4)}    \left(x_{2},y_{2},y_{3},y_{4}\right)+
    \left(y_{1}\rightleftarrows y_{2}\right)+
    \left(y_{1}\rightleftarrows y_{3}\right)+
    \left(y_{1}\rightleftarrows y_{4}\right).
    \label{four_particle_function_FRG2}
\end{equation}

For the translation-invariant problem the expressions (\ref{four_particle_function_FRG1})--(\ref{four_particle_function_FRG2}) in momentum representation are (special kinematics $k_{1}+k_{2}+k_{3}+k_{4}=0$ is assumed):
\begin{equation}
    \frac{\delta\mathcal{G}^{(4)}
    \left[\varLambda\right]
    \left(k_{1},k_{2},k_{3},k_{4}\right)}
    {\delta\varLambda\left(z\right)}=
    \mathcal{G}^{(4)}\left[\varLambda\right] \left(k_{1},k_{2},k_{3},k_{4}\right)
    \sum_{j=1}^{4}\frac{\delta G\left[\varLambda\right]
    \left(k_{j}\right)}
    {\delta\varLambda\left(z\right)}\,
    \mathcal{G}^{(2)}\left[\varLambda\right]
    \left(-k_{j}\right).
    \label{four_particle_function_FRG3}
\end{equation}

Further, for the delta-field configuration of $\varLambda$ (\ref{Delta-field_configuration}) and if $z=w$, the expression (\ref{four_particle_function_FRG3}) is converted to the following form (omitting explicit $w$-dependence):
\begin{equation}
    \frac{\partial\mathcal{G}^{(4)}
    \left(\varLambda;k_{1},k_{2},k_{3},k_{4}\right)}
    {\partial\varLambda}=
    \mathcal{G}^{(4)}
    \left(\varLambda;k_{1},k_{2},k_{3},k_{4}\right)
    \sum_{j=1}^{4}\frac{\partial
    G\left(\varLambda;k_{j}\right)}
    {\partial\varLambda}\,
    \mathcal{G}^{(2)}
    \left(\varLambda;-k_{j}\right).
    \label{four_particle_function_FRG4}
\end{equation}

The solution to the equation (\ref{four_particle_function_FRG4}) satisfying the boundary condition (\ref{ch4-eq11}) with the interaction part of initial action $S_{\mathrm{int}}$ from (\ref{ch4-eq16}) reads as follows:
\begin{equation}
    \mathcal{G}^{(4)}
    \left(\varLambda;k_{1},k_{2},k_{3},k_{4}\right)=
    -g\exp{\left\{\sum_{j=1}^{4}
    \int_{\varLambda_{0}}^{\varLambda}
    d\varLambda'\,\frac{\partial
    G\left(\varLambda';k_{j}\right)}
    {\partial\varLambda'}\,
    \mathcal{G}^{(2)}
    \left(\varLambda';-k_{j}\right)\right\}}.
    \label{four_particle_function_FRG5}
\end{equation}

Let us note that the solution (\ref{four_particle_function_FRG5}) depends only on the absolute values of the corresponding vectors $k_{1},\ldots,k_{4}$. For the optimized additive regulator (\ref{3frgstt6}) the two-particle amputated GF $\mathcal{G}^{(2)}\left(\varLambda;k\right)$ is ``frozen'', so the solution (\ref{four_particle_function_FRG5}) is simplified:
\begin{equation}
    \mathcal{G}^{(4)}
    \left(\varLambda;k_{1},k_{2},k_{3},k_{4}\right)=
    -g\exp{\left\{\sum_{j=1}^{4}
    \int_{\varLambda_{0}}^{\max{\left\{k_{j},
    \varLambda\right\}}}
    d\varLambda'\,\frac{\partial
    G\left(\varLambda'\right)}
    {\partial\varLambda'}\,
    \mathcal{G}^{(2)}
    \left(\varLambda'\right)\right\}}.
    \label{four_particle_function_FRG6}
\end{equation}

Let us consider again the case $n=1$, $D=3$ and the limit $\varLambda_{0}\rightarrow\infty$ (this limit exists, because the polynomial $\varphi^{4}$ theory (\ref{ch4-eq16}) is renormalizable). Calculating the integral appearing in the right-hand side of the expression (\ref{four_particle_function_FRG6}) by changing variables $\varLambda\rightarrow\lambda$ and setting $\varLambda=0$, according to the general principles of the FRG method we obtain the physical four-particle amputated GF $\mathcal{G}^{\mathrm{(Quant)}}$: 
\begin{equation}
    \mathcal{G}^{\mathrm{(Quant)}}
    \left(k_{1},k_{2},k_{3},k_{4}\right)=
    -16g\prod_{j=1}^{4}
    \frac{1}{\kappa_{j}\left(\kappa_{j}
    K_{0}\left(\kappa_{j}\right)+
    2K_{1}\left(\kappa_{j}\right)\right)},\quad
    \kappa_{j}=\frac{4\sqrt{g_{3}}}
    {\sqrt{k_{j}}}.
    \label{four_particle_function_FRG7}
\end{equation}

In double-special kinematics $k_{1}=k,\,k_{2}=-k,\,k_{3}=k',\,k_{4}=-k'$, which is often encountered in equations for low-order GFs, as well as in scattering experiments, the expression (\ref{four_particle_function_FRG7}) has the form:
\begin{equation}
    \mathcal{G}^{\mathrm{(Quant)}}
    \left(k,-k,k',-k'\right)=
    \frac{-16g}{\left(\kappa\kappa'\right)^{2}
    \left(\kappa K_{0}\left(\kappa\right)+
    2K_{1}\left(\kappa\right)\right)^{2}
    \left(\kappa' K_{0}\left(\kappa'\right)+
    2K_{1}\left(\kappa'\right)\right)^{2}},\quad
    \kappa'=\frac{4\sqrt{g_{3}}}
    {\sqrt{k'}}.
    \label{four_particle_function_FRG8}
\end{equation}

At the end of this subsection, let us note one more important advantage of double-special kinematics: This kinematics is required for the iterative solution of the equation for the two-particle amputated GF. Let us also note that the solutions for GFs in terms of Bessel functions are standard solutions in the HRG method \cite{lizana2016holographic,akhmedov1998remark,de2000holographic,verlinde2000rg,fukuma2003holographic,akhmedov2003notes,akhmedov2011hints,heemskerk2011holographic}.

\subsection{Modes Coarse Graining Growth Functionals Rigorous Derivation}

In this final subsection, we check the validity of the statement that the modes coarse graining growth functional $\varPsi$ depends only on the functional momentum $\varpi$, and doesn't depend on derivatives of $\varpi$ with respect to $\varphi$, in other words, to what extent the functional $\varPsi$ can be considered ``geometric''. For this reason let us introduce the so-called blocking functional $\mathcal{B}_{\varLambda}\left[\varphi\right]
\left(x\right)$ (the notation of this subsection coincide with the notation from the review \cite{rosten2012fundamentals}). It is important to note that there is no inverse blocking functional $\mathcal{B}^{-1}_{\varLambda}
\left[\varphi\right]\left(x\right)$. For this reason, Wilsonian renormalization group is only a semigroup. Otherwise the following definition would be trivial. The definition of the Wilsonian effective action $\mathcal{G}$ (we use the same notation as for the generating functional of the amputated GFs, and this should not lead to confusion since this subsection is final) in terms of known initial (bare) action $S$ and functional integral with the integration measure $\mathcal{D}\left[\varphi_{0}\right]$ reads as follows \cite{rosten2012fundamentals}:
\begin{equation}
    e^{-\mathcal{G}_{\varepsilon,\varLambda}
    \left[\varphi\right]}=
    \int \mathcal{D}\left[\varphi_{0}\right]\, \delta^{\left(\infty\right)}_{\varepsilon}
    \left[\varphi-\mathcal{B}_{\varLambda}
    \left[\varphi_{0}\right]\right]
    e^{-S\left[\varphi_{0}\right]}.
    \label{WP_generation_functional1}
\end{equation}

The definition of the modes coarse graining growth functional $\varPsi$ in terms of the blocking functional $\mathcal{B}_{\varLambda}\left[\varphi\right]
\left(x\right)$ is formulated using the expression (\ref{WP_generation_functional1}):
\begin{equation}
    \varPsi_{\varepsilon,\varLambda}
    \left[\varphi\right]\left(x\right)
    e^{-\mathcal{G}_{\varepsilon,\varLambda}
    \left[\varphi\right]}=
    \int \mathcal{D}\left[\varphi_{0}\right]\, \delta^{\left(\infty\right)}_{\varepsilon}
    \left[\varphi-\mathcal{B}_{\varLambda}
    \left[\varphi_{0}\right]\right]
    e^{-S\left[\varphi_{0}\right]}
    \varLambda\frac{\partial 
    \mathcal{B}_{\varLambda}
    \left[\varphi_{0}\right]
    \left(x\right)}{\partial\varLambda}.
    \label{WP_generation_functional2}
\end{equation}

The smoothed Dirac delta-functional $\delta^{\left(\infty\right)}_{\varepsilon}
\left[\varphi\right]$ with normalization constant $\delta^{\left(\infty\right)}_{\varepsilon}
\left[0\right]$, where the step is smeared out over an interval of order $\varepsilon$ (we can recover the sharp Dirac delta-functional by
letting the step width shrink to zero):
\begin{equation}
    \delta^{\left(\infty\right)}_{\varepsilon}
    \left[\varphi\right]=
    \delta^{\left(\infty\right)}_{\varepsilon}
    \left[0\right]
    e^{-\frac{1}{2\varepsilon}
    \left(\varphi|\varphi\right)}.
    \label{WP_generation_functional3}
\end{equation}

Further, consider the master relation for the derivation of the functional $\varPsi$ in terms of the functional $\mathcal{G}$ ($\mathcal{F}$ is an arbitrary functional). This relation plays the role of the source trick in deriving the Wilson--Polchinski and Wetterich--Morris FRG flow equations in \cite{kopbarsch}:
\begin{equation}
    \mathcal{F}_{\varLambda}
    \left[\varepsilon\frac{\delta}
    {\delta\varphi}\right]\left(x\right)
    \delta^{\left(\infty\right)}_{\varepsilon}
    \left[\varphi-\mathcal{B}_{\varLambda}
    \left[\varphi_{0}\right]\right]=
    \mathcal{F}_{\varLambda}
    \left[\varphi-\mathcal{B}_{\varLambda}
    \left[\varphi_{0}\right]\right]\left(x\right)
    \delta^{\left(\infty\right)}_{\varepsilon}
    \left[\varphi-\mathcal{B}_{\varLambda}
    \left[\varphi_{0}\right]\right]+
    \varepsilon\,\mathcal{O}
    \left[\dots\right].
    \label{WP_generation_functional4}
\end{equation}

The functional $\varepsilon\,\mathcal{O}
\left[\dots\right]$ is infinitesimal in the limit $\varepsilon\rightarrow 0$. For this reason, in all the subsequent formulas, this functional is omitted. Further, the functional $\mathcal{F}$ can be chosen in an infinite number of ways. As the first way let $\mathcal{F}$ be a \emph{linear} functional with respect to the field $\varphi-\mathcal{B}_{\varLambda}
\left[\varphi_{0}\right]$:
\begin{equation}
\begin{split}
    &\mathcal{F}_{\varLambda}
    \left[\varphi-\mathcal{B}_{\varLambda}
    \left[\varphi_{0}\right]\right]\left(x\right)=
    \mathcal{F}_{\varLambda}
    \left[\varphi\right]\left(x\right)-
    \mathcal{F}_{\varLambda}
    \left[\mathcal{B}_{\varLambda}
    \left[\varphi_{0}\right]\right]
    \left(x\right),\\
    &\mathcal{F}_{\varLambda}
    \left[\varphi\right]\left(x\right)=
    \left(F_{\varLambda}
    \left(x\right)|\varphi\right)=
    \int d^{D}y\,F_{\varLambda}
    \left(x,y\right)\,
    \varphi\left(y\right).
    \label{WP_generation_functional5}
\end{split}
\end{equation}

We require the following important relation to hold (this relation expresses unknown functional $\mathcal{F}$ in terms of known blocking functional $\mathcal{B}_{\varLambda}\left[\varphi\right]
\left(x\right)$ and its derivative with respect to $\varLambda$):
\begin{equation}
    \mathcal{F}_{\varLambda}
    \left[\mathcal{B}_{\varLambda}
    \left[\varphi_{0}\right]\right]\left(x\right)=
    \varLambda\frac{\partial 
    \mathcal{B}_{\varLambda}\left[\varphi_{0}\right]
    \left(x\right)}{\partial\varLambda}.
    \label{WP_generation_functional6}
\end{equation}

In this case, the functional $\varPsi$ in terms of the functional $\mathcal{G}$ reads as follows:
\begin{equation}
    \varPsi_{\varepsilon,\varLambda}
    \left[\varphi\right]
    \left(x\right)=
    e^{\mathcal{G}_{\varepsilon,
    \varLambda}\left[\varphi\right]}
    \mathcal{F}_{\varLambda}
    \left[\varepsilon\frac{\delta}
    {\delta\varphi}\right]
    \left(x\right)e^{-\mathcal{G}_{\varepsilon,
    \varLambda}\left[\varphi\right]}+
    \mathcal{F}_{\varLambda}\left[\varphi\right]
    \left(x\right).
    \label{WP_generation_functional7}
\end{equation}

Since the functional $\mathcal{F}$ is linear, the following relation holds:
\begin{equation}
    \mathcal{F}_{\varLambda}
    \left[\varepsilon\frac{\delta}
    {\delta\varphi}\right]
    \left(x\right)e^{-\mathcal{G}_{\varepsilon,
    \varLambda}\left[\varphi\right]}=
    -e^{-\mathcal{G}_{\varepsilon,\varLambda}
    \left[\varphi\right]}\int d^{D}y\,F_{\varLambda}\left(x,y\right)\,
    \varepsilon
    \frac{\delta\mathcal{G}_{\varepsilon,\varLambda}
    \left[\varphi\right]}
    {\delta\varphi\left(y\right)}.
    \label{WP_generation_functional8}
\end{equation}

Thus, we arrive at the following expression for the functional $\varPsi$ in terms of the functional $\mathcal{G}$:
\begin{equation}
    \varPsi_{\varepsilon,\varLambda}
    \left[\varphi\right]
    \left(x\right)=
    -\int d^{D}y\,
    F_{\varLambda}\left(x,y\right)\,
    \varepsilon
    \frac{\delta\mathcal{G}_{\varepsilon,
    \varLambda}\left[\varphi\right]}{\delta
    \varphi\left(y\right)}+\int d^{D}y\,
    F_{\varLambda}\left(x,y\right)\,
    \varphi\left(y\right).
    \label{WP_generation_functional9}
\end{equation}

The expression (\ref{WP_generation_functional9}) coincides with the choice (\ref{ch4-eq7}) since the linear term in the right-hand side of the expression (\ref{WP_generation_functional9}) can be easily removed by redefining the functional $\mathcal{G}$. Thus, the conjecture that the functional $\varPsi$ can be considered ``geometric'' is proved in the simplest case when $\mathcal{F}$ is a linear functional with respect to the field $\varphi-\mathcal{B}_{\varLambda}
\left[\varphi_{0}\right]$. 

Let us recall that the functional $\varPsi$ is called geometric if it depends only on the first derivative of the functional $\mathcal{G}$ with respect to the field $\varphi$ which is the functional momentum $\varpi$. If we substitute such a functional $\varPsi$ into the functional master equation (\ref{ch4-eq5}) or (\ref{ch4-eq6}) and neglect the quantum part of the equation, such a functional $\varPsi$ leads to the functional HJ equation, possibly containing higher (third and higher) powers of momentum $\varpi$. In classical mechanics HJ equations are studied by the methods of the symplectic geometry and thus have a number of geometric interpretations. For this reason, the functional $\varPsi$ containing only the first derivative of $\mathcal{G}$ is called geometric.

Further, let us derive the functional equation of a WP type that the Wilsonian effective action $\mathcal{G}$ satisfies (for simplicity of the following expressions, we use the compact notation $\varPhi=\varphi-\mathcal{B}_{\varLambda}
\left[\varphi_{0}\right]$). For this purpose, we calculate the following derivatives:
\begin{equation}
    \varepsilon\varLambda\frac{\partial}{\partial
    \varLambda}e^{-\mathcal{G}_{\varepsilon,
    \varLambda}\left[\varphi\right]}=
    \int\mathcal{D}\left[\varphi_{0}
    \right]\,
    \delta^{\left(\infty\right)}_
    {\varepsilon}
    \left[\varPhi\right]
    e^{-S\left[\varphi_{0}\right]}
    \int d^{D}x\,\varPhi\left(x\right)
    \varLambda
    \frac{\partial\mathcal{B}_{\varLambda}
    \left[\varphi_{0}\right]\left(x\right)}
    {\partial\varLambda}.
    \label{WP_generation_functional14}
\end{equation}
\begin{equation}
    \varepsilon
    \frac{\delta}{\delta\varphi\left(x\right)}
    \left\{\varPsi_{\varepsilon, \varLambda}
    \left[\varphi\right]\left(x\right)
    e^{-\mathcal{G}_{\varepsilon,
    \varLambda}\left[\varphi\right]}\right\}=
    -\int\mathcal{D}\left[\varphi_{0}\right]\, \delta^{\left(\infty\right)}_{\varepsilon}
    \left[\varPhi\right]
    e^{-S\left[\varphi_{0}\right]}
    \varPhi\left(x\right)
    \varLambda\frac{\partial \mathcal{B}_{\varLambda}
    \left[\varphi_{0}\right]\left(x\right)}
    {\partial\varLambda}.
    \label{WP_generation_functional15}
\end{equation}

Comparing the expressions (\ref{WP_generation_functional14}) and (\ref{WP_generation_functional15}), we arrive at the following master equation:
\begin{equation}
    -\varepsilon\varLambda\frac{\partial}{\partial
    \varLambda}e^{-\mathcal{G}_{\varepsilon,
    \varLambda}\left[\varphi\right]}=\int d^{D}x\, 
    \varepsilon\frac{\delta}{\delta\varphi
    \left(x\right)}\left\{\varPsi_{\varepsilon, 
    \varLambda}\left[\varphi\right]\left(x\right)
    e^{-\mathcal{G}_{\varepsilon,
    \varLambda}\left[\varphi\right]}
    \right\}.
    \label{WP_generation_functional17}
\end{equation}

The master equation (\ref{WP_generation_functional17}) can be transformed to the following pseudo-linear form:
\begin{equation}
    \varepsilon\varLambda\frac{\partial
    \mathcal{G}_{\varepsilon,\varLambda}
    \left[\varphi\right]}{\partial\varLambda}=
    \int d^{D}x
    \left\{\varepsilon
    \frac{\delta\varPsi_{\varepsilon,
    \varLambda}\left[\varphi\right]\left(
    x\right)}{\delta\varphi\left(x\right)}-
    \varPsi_{\varepsilon,\varLambda}
    \left[\varphi\right]\left(x\right)
    \varepsilon\frac{\delta\mathcal{G}_
    {\varepsilon,\varLambda}
    \left[\varphi\right]}
    {\delta\varphi\left(x\right)}\right\}.
    \label{WP_generation_functional19}
\end{equation}

The master equation (\ref{WP_generation_functional19}) is valid for any functional $\varPsi$ defined in (\ref{WP_generation_functional2}). Thus, the equation (\ref{WP_generation_functional19}) is an implicit equation for the Wilsonian effective action $\mathcal{G}$. Now let us use the expression (\ref{WP_generation_functional9}) in order to obtain an explicit equation for the functional $\mathcal{G}$. This equation reads: 
\begin{equation}
\begin{split}
    &\varepsilon\varLambda\frac{\partial
    \mathcal{G}_{\varepsilon,\varLambda}
    \left[\varphi\right]}{\partial\varLambda}=
    \int d^{D}x \int d^{D}y \mathcal{F}_
    {\varLambda}\left(x,y\right)\bigg\{
    \varepsilon\frac{\delta\mathcal{G}_
    {\varepsilon,\varLambda}\left[
    \varphi\right]}{\delta\varphi\left(
    x\right)}\varepsilon\frac{\delta
    \mathcal{G}_{\varepsilon,\varLambda}
    \left[\varphi\right]}{\delta\varphi
    \left(y\right)}-\\
    &-\varepsilon^{2}\frac
    {\delta^{2}\mathcal{G}_{\varepsilon,
    \varLambda}\left[\varphi\right]}
    {\delta\varphi\left(x\right)\delta
    \varphi\left(y\right)}-\varphi
    \left(y\right)\varepsilon\frac
    {\delta\mathcal{G}_{\varepsilon,
    \varLambda}\left[\varphi\right]}
    {\delta\varphi\left(x\right)}\bigg\}+
    \varepsilon\,\mathcal{O}
    \left[\dots\right].
    \label{WP_generation_functional20}
\end{split}
\end{equation}

The functional $\varepsilon\,\mathcal{O}
\left[\dots\right]$ is infinitesimal in the limit $\varepsilon\rightarrow 0$. For this reason, this term is never used in all the practical calculations. Thus, we have obtained the functional equation of a WP type that the Wilsonian effective action $\mathcal{G}$ satisfies.

Traveling back to the conjecture, let $\mathcal{F}$ be a \emph{quadratic} functional with respect to the field $\varphi-\mathcal{B}_{\varLambda}
\left[\varphi_{0}\right]$. The expression (\ref{WP_generation_functional4}) reads as follows (we use the compact notation $\varPhi=\varphi-\mathcal{B}_{\varLambda}
\left[\varphi_{0}\right]$ again):
\begin{equation}
    \mathcal{F}_{\varLambda}
    \left[\varepsilon\frac{\delta}
    {\delta\varphi}\right]\left(x\right)
    \delta^{(\infty)}_{\varepsilon}
    \left[\varPhi\right]=
    \left(\varepsilon\frac{\delta}
    {\delta\varphi}\bigg|\hat{\mathcal{Q}}_
    {\varLambda}\left(x\right)\bigg|
    \varepsilon\frac{\delta}
    {\delta\varphi}\right)
    \delta^{(\infty)}_{\varepsilon}
    \left[\varPhi\right]=
    \delta^{(\infty)}_{\varepsilon}
    \left[\varPhi\right]
    \left(\varPhi\big|\hat{\mathcal{Q}}_
    {\varLambda}\left(x\right)\big|
    \varPhi\right)+
    \varepsilon\,\mathcal{O}
    \left[\dots\right].
    \label{WP_generation_functional21}
\end{equation}

The quadratic form in the right-hand side of the expression (\ref{WP_generation_functional21}) has the form:
\begin{equation}
    \left(\varPhi\big|\hat{\mathcal{Q}}_
    {\varLambda}\left(x\right)\big|
    \varPhi\right)=\int d^{D}y\int d^{D}z\,Q_{\varLambda}\left(x,y,z\right)\,
    \varPhi\left(y\right)\varPhi\left(z\right).
    \label{WP_generation_functional22}
\end{equation}

Further, the expression (\ref{WP_generation_functional21}) can be transformed as follows:
\begin{equation}
    \left(\varepsilon\frac{\delta}
    {\delta\varphi}+\varphi\bigg|
    \hat{\mathcal{Q}}_
    {\varLambda}\left(x\right)\bigg|
    \varepsilon\frac{\delta}
    {\delta\varphi}+\varphi\right)
    \delta^{(\infty)}_{\varepsilon}
    \left[\varPhi\right]=
    \left(\mathcal{B}_{\varLambda}
    \left[\varphi_{0}\right]\big|
    \hat{\mathcal{Q}}_
    {\varLambda}\left(x\right)
    \big|\mathcal{B}_{\varLambda}
    \left[\varphi_{0}\right]\right)
    \delta^{(\infty)}_{\varepsilon}
    \left[\varPhi\right]+\varepsilon\,
    \mathcal{O}
    \left[\dots\right].
    \label{WP_generation_functional23}
\end{equation}

Taking into account condition (\ref{WP_generation_functional6}), we arrive at the following expression for the functional $\varPsi$ in terms of the functional $\mathcal{G}$:
\begin{equation}
    \varPsi_{\varepsilon,\varLambda}
    \left[\varphi\right]
    \left(x\right)=
    e^{\mathcal{G}_{\varepsilon,
    \varLambda}\left[\varphi\right]}
    \left(\varepsilon\frac{\delta}
    {\delta\varphi}+\varphi\bigg|
    \hat{\mathcal{Q}}_
    {\varLambda}\left(x\right)\bigg|
    \varepsilon\frac{\delta}
    {\delta\varphi}+\varphi\right)
    e^{-\mathcal{G}_{\varepsilon,
    \varLambda}\left[\varphi\right]}+
    \varepsilon\,
    \mathcal{O}
    \left[\dots\right].
    \label{WP_generation_functional26}
\end{equation}

The expression (\ref{WP_generation_functional26}) shows that the functional $\varPsi$ depends not only on the functional momentum $\varpi$, but also on the derivatives of $\varpi$ with respect to $\varphi$. However, according to the idea of the semiclassical approximation, these derivatives should be replaced by derivatives of known initial (bare) action $S$. 
Thus, within the framework of the semiclassical approximation, we can indeed consider the functional $\varPsi$ to be ``geometric''. The validity of this statement is proved.

At the end of the section, we make one interesting remark. As known, HJ equation is not the only formalism in classical mechanics. Alternatively, one can use Hamilton formalism. Of particular interest is a functional generalization of Hamiltonian mechanics (the introduction of the Hamilton functional equations for the ``canonical coordinates'' $\varphi\left[\varLambda\right]$ and $\varpi\left[\varLambda\right]$) and its connection with the functional HJ equation. Also of interest is the question of the role played by the functional generalization of Hamiltonian mechanics in the FRG and HRG methods. Let us also note that such a generalization would be useful for many-particle physics.

\section{Conclusions}

In this paper we have presented a comprehensive study of functional equations that contain variational derivatives in coordinate representation as well as in condensed notation: The functional Hamilton–Jacobi equation, the functional Schr\"{o}dinger equation, and generalized Wilson–Polchinski functional equation. In terms of the holographic field $\varLambda$, we have given the rigorous derivation of the corresponding hierarchies. We have obtained translation-invariant solution of the first equations of the HJ hierarchy for various configurations of $\varLambda$. Further, the obtained pair of hierarchies for quantum (containing the quantum correction to the potential) HJ and continuity functional equations is the starting point for an approximate solution of the functional Schr\"{o}dinger equation. For this equation we have derived the functional semiclassical approximation and found the translation-invariant solution for various GFs. We have also presented the optical potential for the functional Schr\"{o}dinger equation, which, therefore, should describe an open quantum field system.

Further in the paper we have formulated a semiclassics for the generalized WP equation that describes the FRG flow of the amputated GFs generating functional. We have analyzed in detail modes coarse graining growth functionals that allow such an approximation within the definition in terms of the functional integral. Let us note that the ``nature'' of this semiclassics is essentially different than that for the functional Schr\"{o}dinger equation. With the optimized (Litim) regulator we have found translation-invariant solution of the approximated WP hierarchy for two-particle and four-particle amputated GFs. We have presented the discussion for physical ($\varLambda=0$) two-particle and four-particle amputated GFs in various kinematics.

The results of this paper provide a solid basis for studies of Stochastic Theory of Turbulence, Nuclear Physics, Theory of Critical Phenomena, Quantum Theory of Magnetism, Plasma Physics, Theory of Open Quantum Systems, Stochastic Partial Differential Equations Theory, etc. A short list of further possible research directions is as follows: Scalar QFT in curved spacetime with extra dimensions and compactification, $N$-component scalar model, different generating functionals, GFs families and composite operators FRG flow equations, analytical continuation to Minkowski spacetime, Dyson--Schwinger and Schwinger--Tomonaga functional equations and iterative procedure for solution, noncommutative scalar QFT, higher regulators in FRG method, FRG formulation of developed turbulence, and even such an ambitious question as the definition of QFT by the functional Cauchy problem (FRG flow equation and boundary condition) instead of the original functional integral. 

As an illustration of the last statement: The standard WP functional equation can be studied in the local potential approximation (LPA) \cite{kopbarsch,Felder}. In this approximation, the functional equation is reduced to the ordinary differential equation of a quantum harmonic oscillator. The behavior of the solution must be polynomial. This leads to the quantization of the space dimension $D$, which enters the equation as a continuous parameter. This result is presented in the paper \cite{Felder}. For this reason, the following question is valid: Is it possible to construct a functional FRG flow equation such that in the LPA it reduces, for example, to the generalized hypergeometric function ordinary differential equation? Moreover, the exact functional equation shouldn't destroy the solution obtained in the LPA. In conclusion, let us note that this illustration doesn't pretend to be rigorous, but it certainly gives some ``food for thought''.

\authorcontributions{Mikhail G. Ivanov: conceptualization, methodology, formal analysis, investigation, writing--original draft preparation, writing--review and editing; Alexey E. Kalugin: software, validation, formal analysis, investigation, writing--original draft preparation, visualization; Anna A. Ogarkova: software, validation, formal analysis, investigation, writing--original draft preparation, visualization; Stanislav L. Ogarkov: conceptualization, methodology, formal analysis, investigation, writing--original draft preparation, writing--review and editing, supervision.}

\funding{This research received no external funding}

\acknowledgments{The authors are deeply grateful to Ivan V. Chebotarev for his help with Overleaf, Online LaTeX Editor. We are very grateful to our colleagues at the Department of Higher Mathematics MIPT and at the Department of Theoretical Physics MIPT for helpful discussions and comments. We express special gratitude to Sergey E. Kuratov and Alexander V. Andriyash for supporting this research at an early stage at the Center for Fundamental and Applied Research (Dukhov Research Institute of Automatics). Finally, we are very grateful to Reviewer 2 for many valuable comments and advice on this paper.}

\conflictsofinterest{The authors declare no conflict of interest.} 

\abbreviations{The following abbreviations are used in this manuscript:\\
\noindent 
\begin{tabular}{@{}ll}
AdS & Anti-de Sitter\\ 
CFT & Conformal Field Theory\\
NL & Newton--Leibniz\\
HJ & Hamilton--Jacobi\\
WP & Wilson--Polchinski\\
GF & Green Function
\end{tabular}
\begin{tabular}{@{}ll}
QM & Quantum Mechanics\\
QFT & Quantum Field Theory\\
RG & Renormalization Group\\
LPA & Local Potential Approximation\\
FRG & Functional Renormalization Group\\
HRG & Holographic Renormalization Group
\end{tabular}}

\newpage

\reftitle{References}

\end{document}